\RequirePackage{fix-cm}
\documentclass[twocolumn,epjc3]{svjour3}        

\RequirePackage[T1]{fontenc}
\smartqed 
\RequirePackage{graphicx}
\RequirePackage{mathptmx}     
\RequirePackage{subfigure}
\RequirePackage{mathrsfs}
\RequirePackage[numbers,sort&compress]{natbib}
\RequirePackage{lineno}

\usepackage{xspace}
\usepackage{atlasphysics}
\usepackage{hyperref}
\usepackage{helvet}
\usepackage{amsmath}
\usepackage{amssymb}

\usepackage{preprintcover}  
\PreprintCoverPaperTitle{Measurements of top quark pair relative differential cross-sections with ATLAS in $pp$ collisions at $\boldmath{\sqrt{s}}$ = 7 \tev} 
\PreprintIdNumber{CERN-PH-EP-2012-165} 
\PreprintCoverAbstract{Measurements are presented of differential cross-sections for top quark pair
production in $pp$ collisions at $\sqrt{s}$ = 7 \tev\  relative to the
total inclusive top quark pair production cross-section.
A data sample of \lumitot\ recorded by the \mbox{ATLAS} detector
at the Large Hadron Collider is used.
Relative differential cross-sections are derived as a
function of the invariant mass, the transverse momentum and the
rapidity of the  top quark pair system. 
Events are selected in the lepton (electron or muon) $+$ jets channel.
The background-subtracted differential distributions are corrected for
detector effects, normalized to the total inclusive top quark pair
production cross-section and compared to theoretical predictions. 
The measurement uncertainties range typically between 10\% and 20\%
and are generally dominated by systematic effects.
No significant deviations from the Standard Model expectations are observed.} 
\PreprintJournalName{Eur. Phys. J. C} 

\journalname{Eur. Phys. J. C}

\newcommand{\jimmy}{{\sc JIMMY}}
\newcommand{\herwig}{{\sc HERWIG}}
\newcommand{\powheg}{{\sc POWHEG}}
\newcommand{\pythia}{{\sc PYTHIA}}
\newcommand{\alpgen}{{\sc ALP\-GEN}}

\newcommand{\mcnlo}{{\sc MC\-@NLO}}
\newcommand{\mcnlonobreak}{{\sc MC@NLO}}
\newcommand{\mcfm}{{\sc MCFM}}

\newcommand{\acermc}{{\sc AcerMC}}
\newcommand{\mttbar}{\ensuremath{m_{t\bar{t}}}}
\newcommand{\ppbar}{\ensuremath{p\bar{p}}}

\newcommand{\muplus} {\ensuremath{\mu+\mathrm{jets}}\xspace}
\newcommand{\eplus} {\ensuremath{e+\mathrm{jets}}\xspace}
\newcommand{\lplus} {\ensuremath{\ell+\mathrm{jets}}\xspace}

\newcommand{\wjets}{\ensuremath{W\!+\!\mathrm{jets}}}
\newcommand{\zjets}{\ensuremath{Z\!+\!\mathrm{jets}}}
\newcommand{\wlightjets}{\ensuremath{W\!+\!\textrm{light-jets}}}
\newcommand{\wccjets}{\ensuremath{Wc\bar{c}\!+\!\mathrm{jets}}}
\newcommand{\wbbjets}{\ensuremath{Wb\bar{b}\!+\!\mathrm{jets}}}
\newcommand{\wcjets}{\ensuremath{Wc\!+\!\mathrm{jets}}}
\newcommand{\wtwojets}{\ensuremath{W\!+\!\textrm{2-jets}}}
\newcommand{\wonejets}{\ensuremath{W\!+\!\textrm{1-jet}}}
\newcommand{\wplusjets}{\ensuremath{W^{+}\!+\!\mathrm{jets}}}
\newcommand{\wminusjets}{\ensuremath{W^{-}\!+\!\mathrm{jets}}}

\newcommand{\mtw}{\ensuremath{m_{\mathrm{T}}^{W}}}

\newcommand{\lumitot}{\mbox{2.05\,fb$^{-1}$}}
\newcommand{\invfb}{\mbox{fb$^{-1}$}}

\begin{document}

\title{Measurements of top quark pair relative differential cross-sections
with ATLAS in $\bold{pp}$ collisions at $\bold{\sqrt{s}}$ = 7 \tev}

\author{ The ATLAS Collaboration\thanksref{e1,addr1}\\
}

\thankstext{e1}{e-mail: atlas.publications@cern.ch}

\institute{CERN, 1211 Geneva 23, Switzerland \label{addr1}
}

\date{Received: date / Accepted: date}

\maketitle

\begin{abstract}
Measurements are presented of differential cross-sections for top quark pair
production in $pp$ collisions at $\sqrt{s}$ = 7 \tev\  relative to the
total inclusive top quark pair production cross-section.
A data sample of \lumitot\ recorded by the \mbox{ATLAS} detector
at the Large Hadron Collider is used.
Relative differential cross-sections are derived as a
function of the invariant mass, the transverse momentum and the
rapidity of the  top quark pair system. 
Events are selected in the lepton (electron or muon) $+$ jets channel.
The background-subtracted differential distributions are corrected for
detector effects, normalized to the total inclusive top quark pair
production cross-section and compared to theoretical predictions. 
The measurement uncertainties range typically between 10\% and 20\%
and are generally dominated by systematic effects.
No significant deviations from the Standard Model expectations are observed.
\end{abstract}

\section{Introduction} 
\label{sec:introduction}

The top quark~\cite{CDFTopDiscAbe:1995hr,D0TopDiscAbachi:1995iq} is
the most massive known fundamental constituent of matter. Its
unexplained large mass suggests an important connection to the
electroweak symmetry breaking mechanism. The measurement of the top-antitop (\ttbar) quark production cross-section
($\sigma_{\ttbar}$) in various decay channels allows a precision test of perturbative QCD.
In addition, the \ttbar\ production process is an important background for
Standard Model (SM)
Higgs boson searches, and in searches for physics beyond the SM.
Also, a rich set of possible new particles and
interactions might appear at the Large Hadron Collider (LHC) and modify the
production and/or decay of top quarks.

The inclusive \ttbar\ production cross-section has
been measured by the ATLAS and CMS Collaborations with increasing
precision~\cite{Aad:2012a,Aad:2012b,Chatrchyan:2011a,Chatrchyan:2011b}
in a variety of channels using
data collected in 2010 and 2011.
The unprecedented number of available \ttbar\ events
(tens of thousands) enables detailed
investigations of the properties of top quark production in terms
of characteristic variables of the \ttbar\ system.
This paper focuses on three observables of the \ttbar\ system: the
invariant mass ($m_{\ttbar}$), the transverse momentum
($p_{\mathrm{T},\ttbar}$) and the rapidity ($y_{\ttbar}$).
To enable direct comparisons to theoretical models the differential
distributions are unfolded for detector effects and corrected for
acceptance effects.
Theoretical predictions for the \ttbar\ invariant mass distribution
accurate to next-to-next-to-leading logarithm (NNLL) and
next-to-leading order (NLO) are currently available~\cite{AHR-0901},
with a typical uncertainty of around 12\% at $m_{\ttbar}$ $\simeq$ 1 \tev.
Comparisons of mass, transverse momentum, and rapidity distributions are
also made between unfolded data and NLO
predictions taken from the MCFM generator~\cite{MCFM}.
In addition, the  data are compared to predictions from the
\mcnlo{}~\cite{FRI-0201,Frixione:2003ei} and \alpgen{}~\cite{MAN-0301} generators with
particular choices of parameter settings.

The $m_{\ttbar}$  distribution is sensitive to particles beyond the SM, such as new {\it s}-channel resonances that
can modify the shape of the differential production cross-section in different
ways depending on their spin and colour properties~\cite{FRED-0901}.
In addition to Tevatron experiment
searches~\cite{CDFLimits2007ag,CDFSearch2007ak,D0Search2008ny,CDFSearch2009tx,CDFSearch2011ts,D0Search2011gv},
both the ATLAS and CMS Collaborations have performed direct searches for specific narrow
and wide resonances that extend mass limits to the \tev\
region~\cite{Aad:2012wm,Aad2012txa,CMS_TTZ}.
The CDF Collaboration has performed a measurement of the differential
cross-section as a function of \mttbar~\cite{CDF-0901}
using the data collected in proton-antiproton  ($\ppbar$) collisions
at a centre of mass energy (\rts) of 1.96~\tev. The result is consistent
with the SM expectation as predicted by PYTHIA (version 6.216)~\cite{SJO-0601}.
A potentially intriguing deviation from the SM prediction is
observed in the measured forward-back\-ward angular asymmetry between 
$t$ and \tbar\ quarks produced together in \ppbar\ collisions
at the Tevatron~\cite{Aaltonen:2011kc, Abazov:2011rq}, particularly in events with large
\mttbar~\cite{Aaltonen:2011kc}.
Nearly all new physics scenarios that could explain this deviation
should be observable at the LHC as a resonant or non-resonant
enhancement with respect to the SM in \ttbar\ production at large
\mttbar~\cite{SAAV-2011}.

\section{Detector, data and simulation samples}
\label{sec:dataSimAndDet}
The ATLAS detector~\cite{Aad:2008zzm} at the LHC covers nearly the
entire solid angle around the collision point. It consists of
an inner tracking detector (ID) comprising a silicon pixel
detector, a silicon microstrip detector, and a transition
radiation tracker, providing tracking capability within
pseudorapidity~\footnote{ATLAS uses a
right-handed coordinate system with 
its origin at the nominal interaction point in the centre of the
detector and the $z$-axis along the beam pipe. 
The $x$-axis points to the centre of the LHC ring, and the
$y$-axis points upward. Cylindrical 
coordinates $(r,\phi)$ are used in the transverse plane, $\phi$ being
the azimuthal angle around the beam pipe. 
The pseudorapidity is defined in terms of the polar angle $\theta$
as $\eta=-\ln\tan(\theta/2)$.
Transverse momentum and energy are defined as $\pt = p\sin\theta$ and
$\ET = E\sin\theta$, respectively. The distance $\Delta R$ is defined as
$\Delta R = \sqrt{({\Delta \phi})^2+{(\Delta \eta})^2}$, where $\Delta\phi$
and $\Delta\eta$ are the separation in azimuthal angle and pseudorapidity,
respectively.} $|\eta| < 2.5$.
The ID is surrounded by a thin superconducting
solenoid providing a 2 T axial magnetic field, and by liquid argon
(LAr) electromagnetic (EM) sampling calorimeters with
high granularity. An iron/scintillator tile calorimeter provides
hadronic energy measurements in the central
pseudorapidity
range ($|\eta| < 1.7$). The end-cap and forward regions are instrumented
with LAr calorimeters for both electromagnetic
and hadronic energy measurements up to $|\eta| < 4.9$. The
calorimeter system is surrounded by a muon spectrometer
incorporating three superconducting toroid magnet assemblies.

A three-level trigger system is used to select high-\pt\ events. The level-1
trigger is implemented
in hardware and uses a subset of the detector information
to reduce the rate to a design value of at most
75 kHz. This is followed by two software-based trigger levels,
which together reduce the event rate to about 300 Hz.
This analysis uses LHC proton-proton ($pp$) collisions at \rts\ = 7~\tev\ collected by
the ATLAS detector between March and August 2011, corresponding to an
integrated luminosity of \lumitot.

Simulated top quark pair events are generated using the
\mcnlo{} Monte Carlo (MC) generator version 3.41 with the NLO parton
distribution function (PDF) set CTEQ6.6~\cite{Nadolsky:2008zw}, where the
top quark mass is set to 172.5 \gev. Renormalization and factorization
  scales are set to the same value: the square root of the average of the $t$ and \tbar\
  quarks squared transverse energies. Parton showering and the underlying event are modelled using
\herwig~\cite{COR-0001} and  \jimmy~\cite{JButterworth:1996zw} using
the AUET1 tune ~\cite{atlastune}, respectively.
The \ttbar\ sample is normalized to a cross-section of 164.6~pb,
obtained with an approximate NNLO prediction~\cite{Aliev:2010zk}.
Single top events are also generated using \mcnlo~\cite{Frixione:2005vw,Frixione:2008yi}, while the production
of $W/Z$ bosons in association with jets is simulated using 
the \alpgen \ generator interfaced to \herwig \ and \jimmy\
with CTEQ6L1 PDFs~\cite{cteq6}. \wjets\ events containing $b\bar{b}$
  pairs, $c\bar{c}$ pairs and single $c$-quark (heavy flavour) were
  generated separately using matrix elements with massive
$b$- and $c$-quarks. An overlap-removal procedure is used to avoid
double counting due to heavy quarks from the parton shower.
Diboson events ($WW$, $WZ$, $ZZ$) are
generated using \herwig\ with MRST $\rm LO^{*}$ PDFs~\cite{Sherstnev:2007nd}.

All Monte Carlo simulation samples are generated with multiple $pp$
interactions per bunch crossing (pile-up).
These simulated events are re-weighted so that the distribution of
the average number of interactions per $pp$ bunch crossing 
in simulation matches that observed in the data.
This average number varies between data-taking periods and ranges
typically between 4 and 8.
The samples are then processed through the GEANT4~\cite{AGO-0301} 
simulation of the ATLAS detector~\cite{2010EPJC...70..823A}  and the
standard ATLAS reconstruction software.

\section{Event selection}
\label{sec:objectAndSel}
Events are selected in the lepton (electron or muon) $+$ jets channel.
The reconstruction of \ttbar\ events in the detector is based on the
identification and reconstruction of electrons, muons, jets and
missing transverse momentum.  
The definitions of these objects are identical to those used
in Ref.~\cite{CHARASY-012}.
The same event selection as in Ref.~\cite{CHARASY-012} is used with the
addition of a requirement on the kinematic likelihood resulting from the event
reconstruction described in Sect.~\ref{sec:evreco}.

\subsection{Object definitions}

Electron candidates are defined as energy deposits in the EM calorimeter associated with well-re\-con\-struct\-ed
tracks of charged particles in the
ID.  The candidates are required to meet stringent identification
criteria based on EM shower shape information, track quality variables
and information from the transition radiation tracker~\cite{Aad:2011mk}.
All candidates are required to have $\et>25$ \gev\ and
$|\eta_{\rm clu}|<2.47$, where $\eta_{\rm clu}$ is the pseudorapidity
of the EM calorimeter cluster associated with the
electron. Candidates in the  transition region between the barrel and
end-cap calorimeters $1.37<|\eta_{\rm clu}|<1.52$ are rejected. 

Muon candidates are reconstructed by combining track segments
in different layers of the muon chambers. Such segments are assembled 
starting from the outermost layer, with a procedure that takes
material effects into account, and are then matched with tracks found
in the ID.  The candidates are then re-fitted, 
exploiting the full track information from both the muon spectrometer
and the ID, and are required to have $\pt>20$ \gev\ and $|\eta|<2.5$. 

Jets are reconstructed with the anti-$k_{t}$ algorithm ~\cite{Cacciari:2008gp}
with a distance parameter of $0.4$ using  clusters formed from
calorimeter cells with significant energy deposits ("topoclusters") at
the EM scale.
The jet energy is then corrected to the hadronic scale
using $\pt$- and $\eta$-dependent correction factors derived
from simulation and validated with data~\cite{Aad:2011he}.

The missing transverse momentum  and its magnitude \met\ are derived from
topoclusters at the EM scale and corrected on the basis of the energy scale of the associated
physics object, if any~\cite{Aad:2012re}. Contributions from
muons are included using their momentum measured from the tracking
and muon spectrometer systems.
The remaining clusters not associated with high-\pt\ objects
are added at the EM scale.

Both  the electron and muon candidates are required to be isolated  to
reduce the backgrounds from hadrons mimicking lepton signatures
and leptons from heavy-flavour decays. 
For electron candidates, the total transverse energy deposited in the
calorimeter in a cone of $\Delta R=0.2$ around the electron
candidate is required to be less than $3.5$ \gev\, after correcting for the
energy associated with the electron and for energy deposited by 
pile-up.  For muon candidates, the isolation is defined in a cone of $\Delta
R=0.3$ around the muon direction. In that region both the sum of track
transverse momenta for tracks with $\pt>1$~\gev\ and the total energy
deposited in the calorimeter are required to be less than 4~\gev,  after
subtracting the contributions from the muon itself.

Jets within $\Delta R=0.2$ of an electron candidate are removed  to
avoid double counting electrons as jets.  Subsequently, muons within
$\Delta R=0.4$ of the centre of a jet with $\pt>20$ \gev\ are removed
in order to reduce the contamination caused by muons from hadron
decays. 

The reconstruction of \ttbar\ events is aided by the ability to 
tag jets from the hadronization of $b$-quarks  using the combination
of two $b$-tagging algorithms~\cite{CombNN}.
One $b$-tagger derives the properties of vertices related to $b$- and
$c$-hadron decays inside jets by assuming the vertices to lie on a line
connecting them to the primary vertex~\footnote{A primary vertex is defined
as a vertex reconstructed  from
a number of high-quality tracks such that the vertex is
spatially compatible with the luminous region of interaction. Primary
vertices in an event are ordered by
$\Sigma_{\mathrm{trk}}$ $p_{\mathrm{T},\mathrm{trk}}^{2}$,
where $p_{\mathrm{T},\mathrm{trk}}$ is the transverse momentum of an
associated track.}.
 A likelihood discriminant between $b$-, $c$-
and light-quark jets is derived by using the number,  the masses, the
track energy fraction, the flight-length significances and  the track
multiplicities of the reconstructed vertices as inputs.
The other $b$-tagging algorithm employs the transverse and longitudinal
impact parameter significances of each track within the jet to derive a 
likelihood  that the jet originates from a $b$-quark. The
results of the two taggers are combined, using a neural network, into a single
discriminating variable. 
The combined tagger operating point chosen for the present
analysis corresponds to a 
70\% tagging efficiency for $b$-jets in simulated $t\bar{t}$
events, while light-flavour jets ($c$-jets) are suppressed by approximately
a factor of 100 (5).

\subsection{Selection of \ttbar\ candidates}\label{evselection}
The lepton $+$ jets channel selection requires the
appropriate single-electron or single-muon trigger to have fired (with
thresholds at 20 \gev{} and 18 \gev{} respectively).
Events passing the trigger selection are required to contain
exactly one reconstructed electron (muon)
with $\et > 25$ \gev\  ($\pt > 20$ \gev).
The events are required to have at least one reconstructed primary
vertex. The primary vertex, corresponding to that with highest $\Sigma_{\mathrm{trk}}$
$p_{\mathrm{T},\mathrm{trk}}^{2}$ is required to be reconstructed from at
least five high-quality tracks.
Jet quality criteria are applied to the data and events are discarded if
any jet with $\pt >$ 20 \gev\ is identified to be due to calorimeter noise
or activity out of time with respect to the LHC beam
crossings~\cite{Aad:2011he}.
The \met\ is required to be larger than 20  (35) \gev\ in the \muplus (\eplus)
channel. The $W$ boson transverse mass (\mtw), derived from the lepton
transverse momentum and the \met{}~\cite{Aad:2010ey}, is required to be larger than 60
\gev\ -- \met{} (25 \gev{}) in the \muplus (\eplus) channel.
 The requirements for the \eplus channel is more stringent in order to reduce the larger fake-lepton
background.
Events are required to have at least four jets with $\pt>25$ \gev\ and
$|\eta|<2.5$, where at least one of these jets is required to be
$b$-tagged. 
Finally, events are retained only if they have a kinematic
likelihood $\ln(L) > -52$ resulting from the event reconstruction described in
Sect.~\ref{sec:evreco} .

\section{Background determination}
\label{sec:bkg}
The main expected backgrounds in the lepton $+$ jets channel are
\wjets\,  which can give rise to the same final state as the \ttbar\ signal,
and fake leptons. They are both estimated using auxiliary measurements. 
The other backgrounds are of electroweak origin and are estimated from
simulation.  All background determination methods are identical to those used
in Ref.~\cite{CHARASY-012}.

\subsection{Fake-lepton background}
\label{fakes}
The multijet background with misidentified and non-prompt leptons
(referred to collectively as fake leptons)
in both the \eplus and \muplus channels is evaluated with
a matrix method, which relies on defining loose and tight
lepton samples~\cite{Aad:2010ey,Aad:2012a}
and measuring the fractions of real ($\epsilon_\mathrm{real}$)
and fake ($\epsilon_\mathrm{fake}$) loose
leptons that are selected as tight leptons.
The fraction $\epsilon_\mathrm{real}$ is measured using data
control samples of $Z$ boson decays to two leptons,
while  $\epsilon_\mathrm{fake}$ is measured from data control
regions dominated by the contributions of fake leptons.
Contributions from \wjets\ and \zjets\ backgrounds
are subtracted in the control regions using Monte Carlo simulation.

For the \muplus channel, the loose data sample is defined by discarding
the isolation requirements in the default
muon selection.
The fake-muon efficiencies are derived from a
low-\mtw\ control region, $\mtw < 20\,\gev$, 
with an additional requirement $\met + \mtw < 60\,\gev$.
The efficiencies for real and fake muons are parameterized
as a function of muon $|\eta|$ and of the leading jet \pt.

For the \eplus channel, the loose
data sample is defined by selecting events with electrons meeting
looser identification criteria.  The 3.5 \gev\ electron isolation
requirement is loosened to 6 \gev. The fake-electron efficiencies are
determined using a low-$\met$ control region
($5\, \gev < \met < 20\, \gev$). 
The efficiencies for real and fake-electrons are parameterized
as a function of electron $|\eta|$.

\subsection{\wjets\ background estimation}
\label{sec:ChAsymm}

The \wjets\ background estimation consists of three steps.

The first step is to determine the flavour composition of the \wjets\
background in the signal region before $b$-tagging. 
Since the theoretical prediction for heavy flavour fractions in \wjets\
suffers from large uncertainties, a data-driven
approach was developed to constrain these fractions with inputs
from MC simulation.
Samples with a lower jet multiplicity, obtained from the selection
described in Sect.~\ref{evselection}, but requiring exactly one or
two jets instead of four or more jets, are analysed. 

 The numbers
$W^{\rm data}_{i,
 \textrm{pre-tag}}, W^{\rm data}_{i, \rm tagged}$, of $W+i$-jets events
in these samples (with $i=1,2$), before and
after applying the $b$-tagging requirement, are calculated from the
observed events by subtracting the small contributions from other
Standard Model processes --- electroweak ($WW$,
$WZ$, $ZZ$, and \zjets) and top (\ttbar\ and single top) processes ---
predicted by the simulation and
by subtracting the fake-lepton background obtained as described in
Sect.~\ref{fakes}.

A system of three equations --- expressing the number of \wonejets{} events
after $b$-tagging and \wtwojets{} events before and
after $b$-tagging --- can be written
with eight independent flavour fractions as the unknowns,
corresponding to fractions of \wbbjets, \wccjets, \wcjets\ and \wlightjets\
events in the one- and two- jet bins before $b$-tagging. In the equations
involving tagged events, the simulation prediction is used to include the
eight tagging probabilities of the different \wjets\ event types.
For each flavour, the fractions in the one-jet and two-jet bins are related
using the simulation's prediction of their ratio. 
These predictions reduce the number of independent fractions to four.
Finally, the ratio of the \wccjets\ to the \wbbjets\ fractions
in the two-jet bin is fixed to the value obtained from simulated
events in order to obtain three independent fractions in the three equations.

The resulting scale factors for the heavy flavour fractions in
simulated \wjets\ events are $1.63\pm 0.76$ for
\wbbjets\ and \wccjets\ events and  $1.11\pm 0.35$ for \wcjets\ events.
These are applied to the relevant Monte Carlo samples.
The uncertanties on these scale factors are derived from systematic
variations of the inputs to the method (see Sect.~\ref{sec:SystematicsMod}).
The fraction of \wlightjets\ events is scaled by a factor 0.83 to keep
the total number of pre-tagged Monte Carlo \wjets\ events fixed.
When applied to the signal region, an additional 25\% uncertainty
is applied to these fractions, corresponding to the
uncertainty in the Monte Carlo prediction for the ratio of flavour
fractions in different jet multiplicities. 

The second step is to determine the overall normalization of
\wjets\ background in events with four or more jets before
$b$-tagging. At the LHC the rate of \wplusjets\ events is larger than that
of \wminusjets\ events because there are more up-type valence quarks in
the proton than down-type valence quarks.
The ratio of \wplusjets\ to \wminusjets\ cross-sections is
predicted much more precisely than the total \wjets\ cross-section
\cite{MSTW2008b,KOM-1001,BER-2011}.
This asymmetry is used to measure 
the total \wjets\ background from the data.
To a good approximation, processes other than \wjets\ give
equal numbers of positively and negatively charged leptons.
Consequently  the total number of \wjets\ events in the selected
sample can be estimated as

\begin{equation}
  \label{chargeaform}
  W_{\ge4, \textrm{pre-tag}}  = N_{W^+} + N_{W^-} = \left( {r_{\rm MC}+1 \over
    r_{\rm MC}-1} \right)
(D^+ - D^-).
\end{equation}
The charge-asymmetric single top contribution is estimated  from
simulation and subtracted.
The values $D^+ (D^-)$ are the total numbers of events in data meeting
the selection criteria described in Sect.~\ref{evselection}, before the
$b$-tagging and
likelihood requirement, with positively (negatively) charged leptons.
The value of  $r_{\rm MC} \equiv {N(pp\rightarrow W^{+} + X ) \over
  N(pp\rightarrow W^{-} + X )}$  is derived from Monte Carlo simulation,
using the same event selection. 
The ratio $r_{\rm MC}$ is  $1.56 \pm 0.06$ in the \eplus channel and
$1.65\pm 0.08$ in the \muplus channel. The largest uncertainties on
$r_{\rm MC}$ derive from uncertainties in  PDFs, the jet energy scale,
and the heavy flavour fractions in \wjets\ events.

Finally, in the third step, the number of \wjets\ events passing
the selection with at least one $b$-tagged jet is determined to
be~\cite{Aad:2010ey}
\begin{equation}
  \label{eq:taggedfrac}
  W_{\ge4, \rm tagged} = W_{\ge4, \textrm{pre-tag}} \cdot f_{2,\rm tagged}
  \cdot k_{2\to\ge4}.
\end{equation}
\noindent 
The value $f_{\rm 2,tagged}\equiv W^\textrm{data}_\textrm{2,
  tagged}/$$W^\textrm{data}_\textrm{2, pre-tag}$
is the fraction of $W+$ 2~jets events meeting the requirement of
having at least one $b$-tagged jet, and
$k_{2\to\ge4}\equiv f^{\rm MC}_{\ge4\rm ,tagged}/$$f^{\rm MC}_{\rm 2,tagged}$
is the ratio of the fractions of simulated \wjets\ events passing the 
requirement of at least one $b$-tagged jet, for at least four and exactly two
jets, respectively.
The value of $f_{2,\rm tagged}$ is found
to be $0.063 \pm 0.005$ in the \eplus channel and $0.068 \pm 0.005$ in
the \muplus channel.
The ratio $k_{2\to\ge4}$ is found to be $2.52\pm 0.36$ in the \eplus
channel and $2.35\pm 0.34$ in the \muplus channel.
The uncertainties include both systematic contributions and
contributions arising from the limited number of simulated events.

\subsection{Other backgrounds}
The numbers of background events from single top production,
\zjets\ and diboson events are evaluated using Monte Carlo simulation.
The prediction for \zjets\ events are normalized to the approximate
NNLO cross-sections as determined by the FEWZ
program~\cite{FEWZ-06}, using the MSTW2008NLO
PDFs~\cite{MSTW2008a,MSTW2008b}. The prediction for
diboson events is normalized to the NLO cross-section as determined
by the \mcfm{} program~\cite{MCFM_DIBOS} using the MSTW2008NLO PDFs.
The approximate NNLO cross-section
results from Refs.~\cite{KIDO-011,KIDO-010a,KIDO-010b} are used to normalize
the predictions for single top events.

\section{Reconstruction}
\label{sec:evreco}

Measurements of differential cross-sections in top quark pair events
require full kinematic reconstruction of the \ttbar\ system. The
reconstruction is performed
using a likelihood fit of the measured objects to a
theoretical leading-order representation of the \ttbar\ decay.
The same reconstruction method as in Ref.~\cite{CHARASY-012} is used.
The likelihood is the product of three factors. The first factor is the
product of Breit-Wigner distributions for the
production of $W$ bosons and top quarks, given the four-momenta of the
true \ttbar\ decay products. 
The second factor is the product of transfer functions representing the
probabilities for the given true
energies of the \ttbar\ decay products to be observed as the energies
of reconstructed jets, leptons and as missing transverse energy. The
third factor is the probability to $b$-tag a certain jet, given the parton
it is associated with.
The pole masses of the $W$ bosons and the top quarks 
in the Breit-Wigner distributions are set to 80.4 GeV and 172.5 GeV, respectively.

\begin{figure*}[htbp]
  \begin{minipage}{\textwidth}
    \begin{center}
      \subfigure[]{
	\includegraphics[scale=0.37]{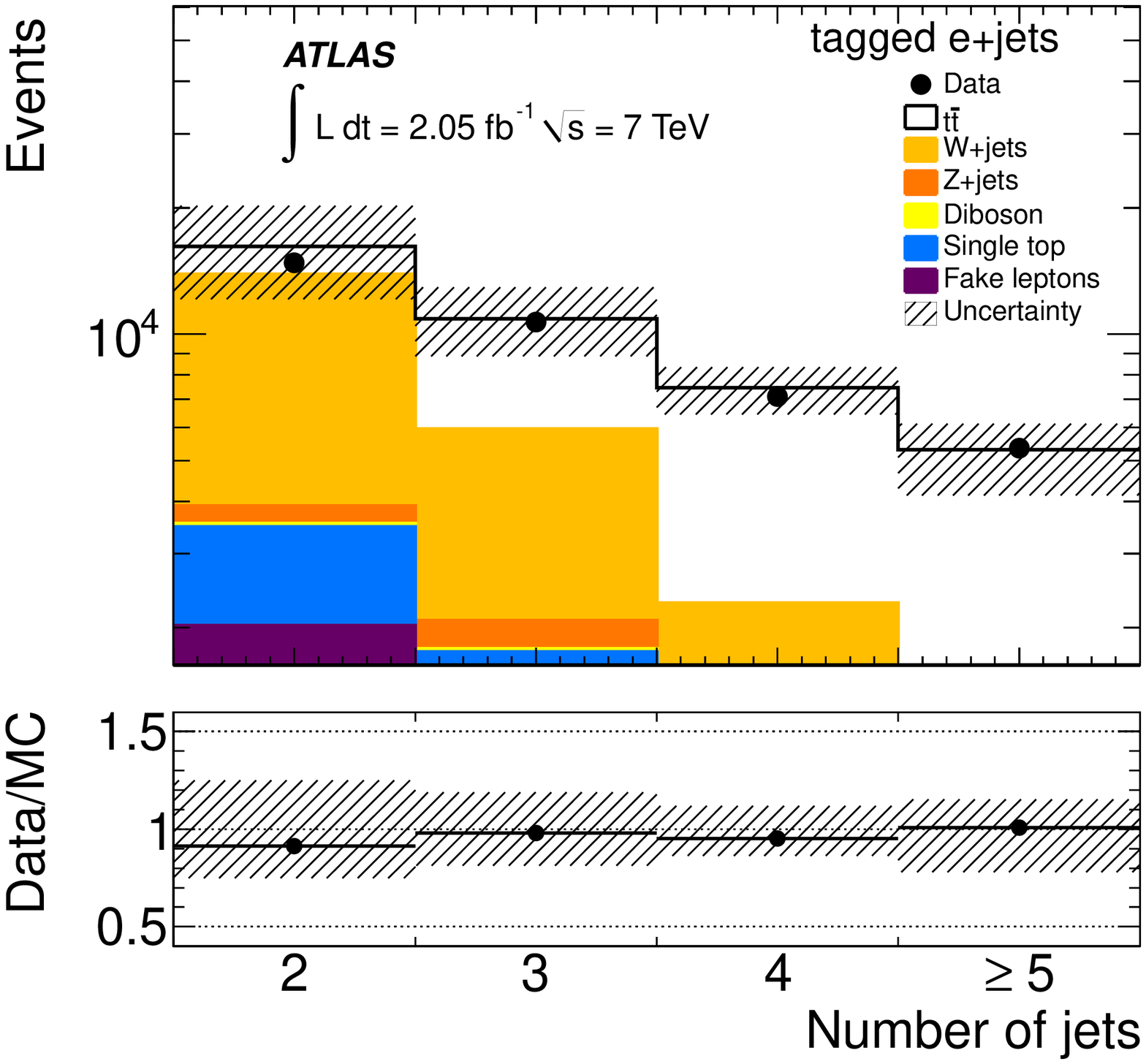}
      }
      \subfigure[]{
	\includegraphics[scale=0.37]{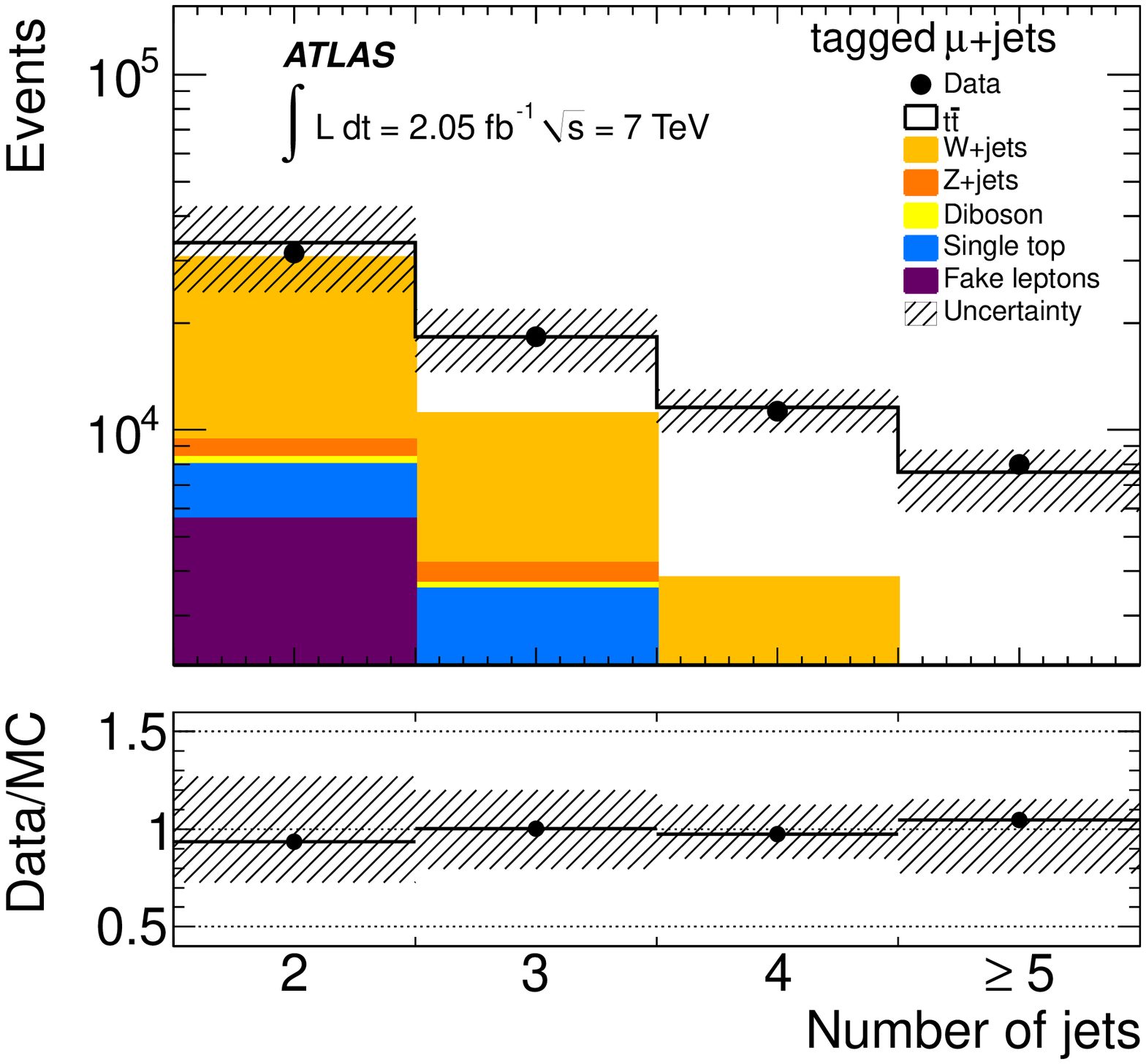}
      }
      \subfigure[]{
        \includegraphics[scale=0.37]{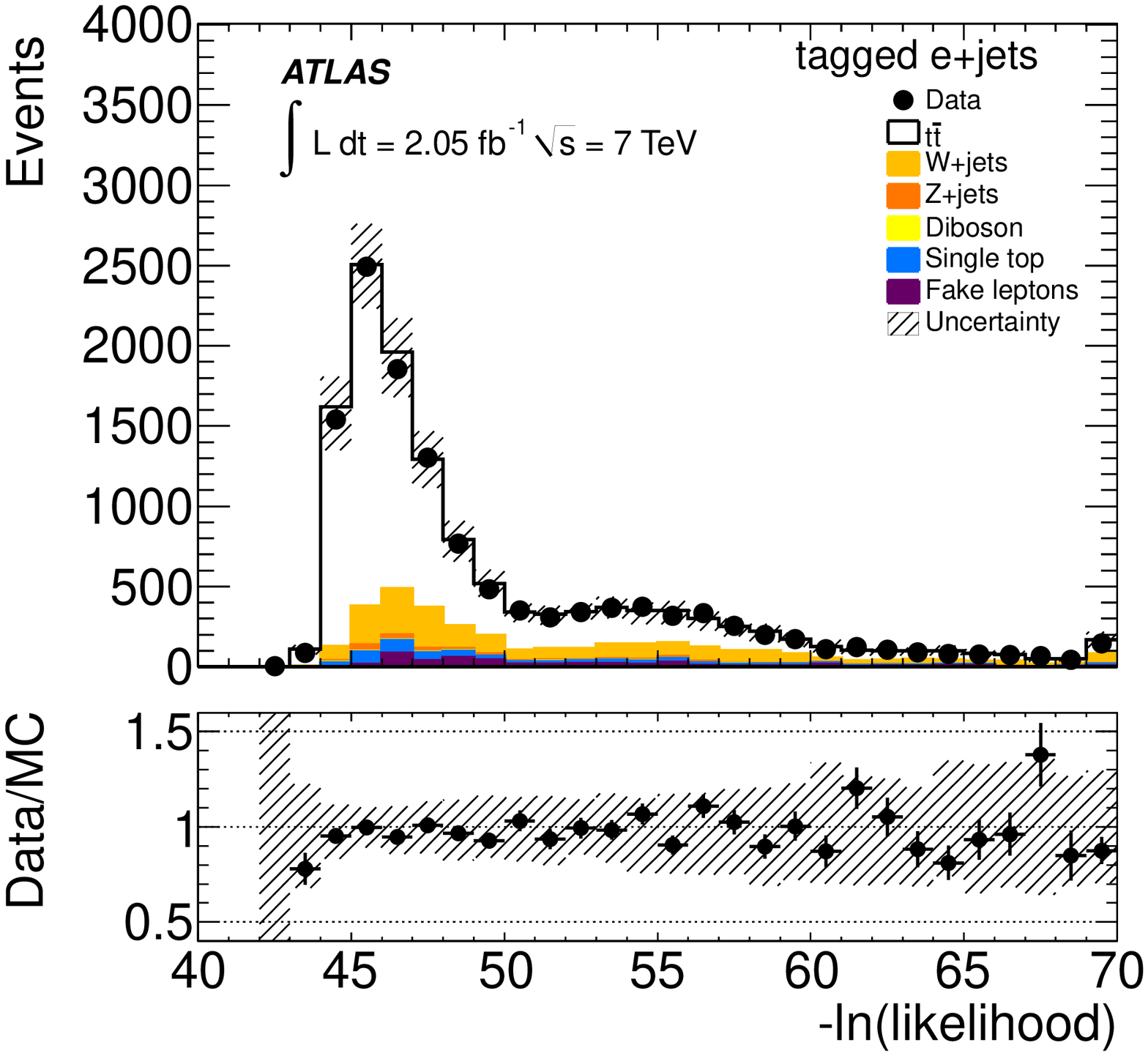}
      }
      \subfigure[]{
        \includegraphics[scale=0.35]{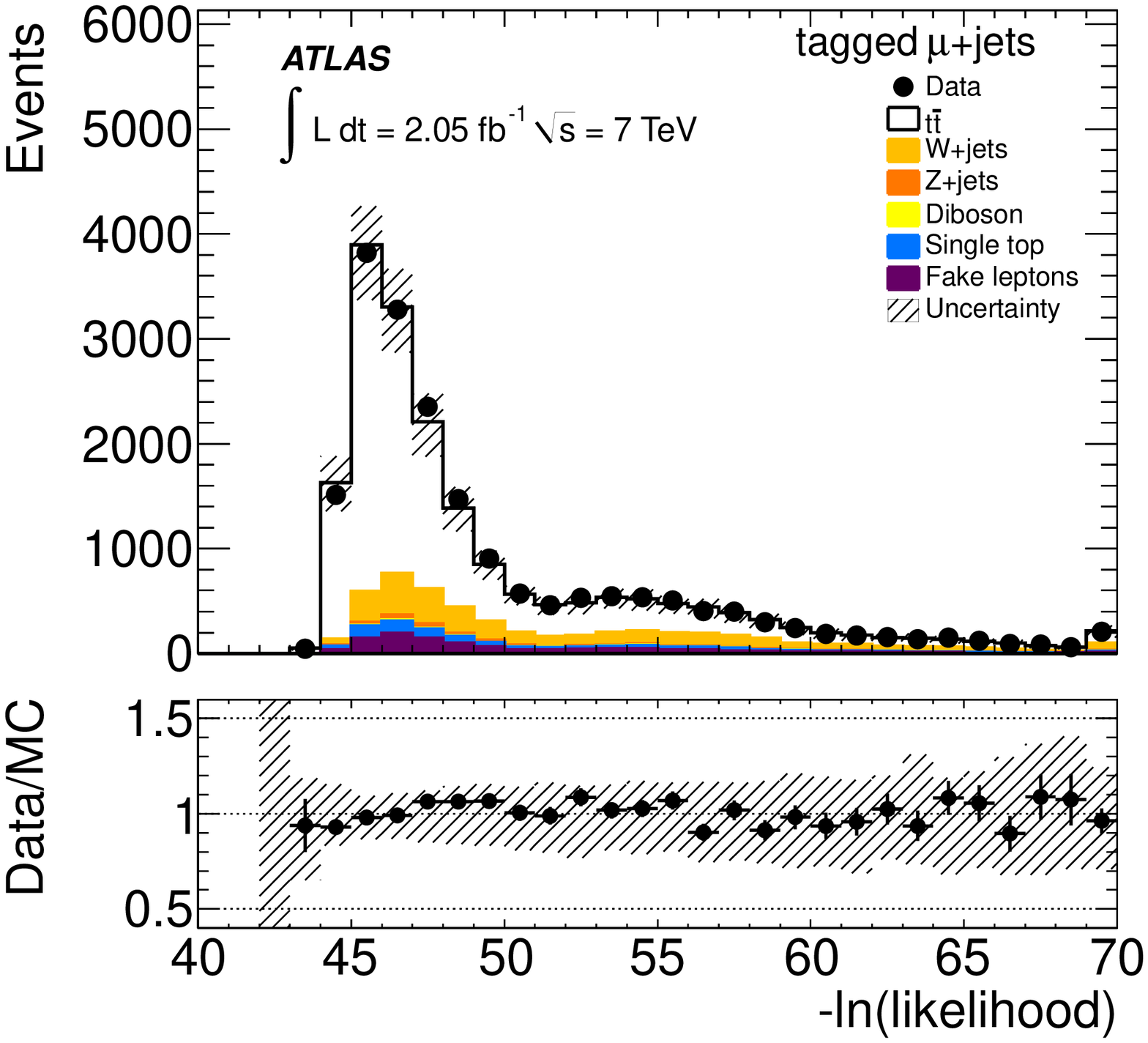}
      }
     \subfigure[]{
	\includegraphics[scale=0.37]{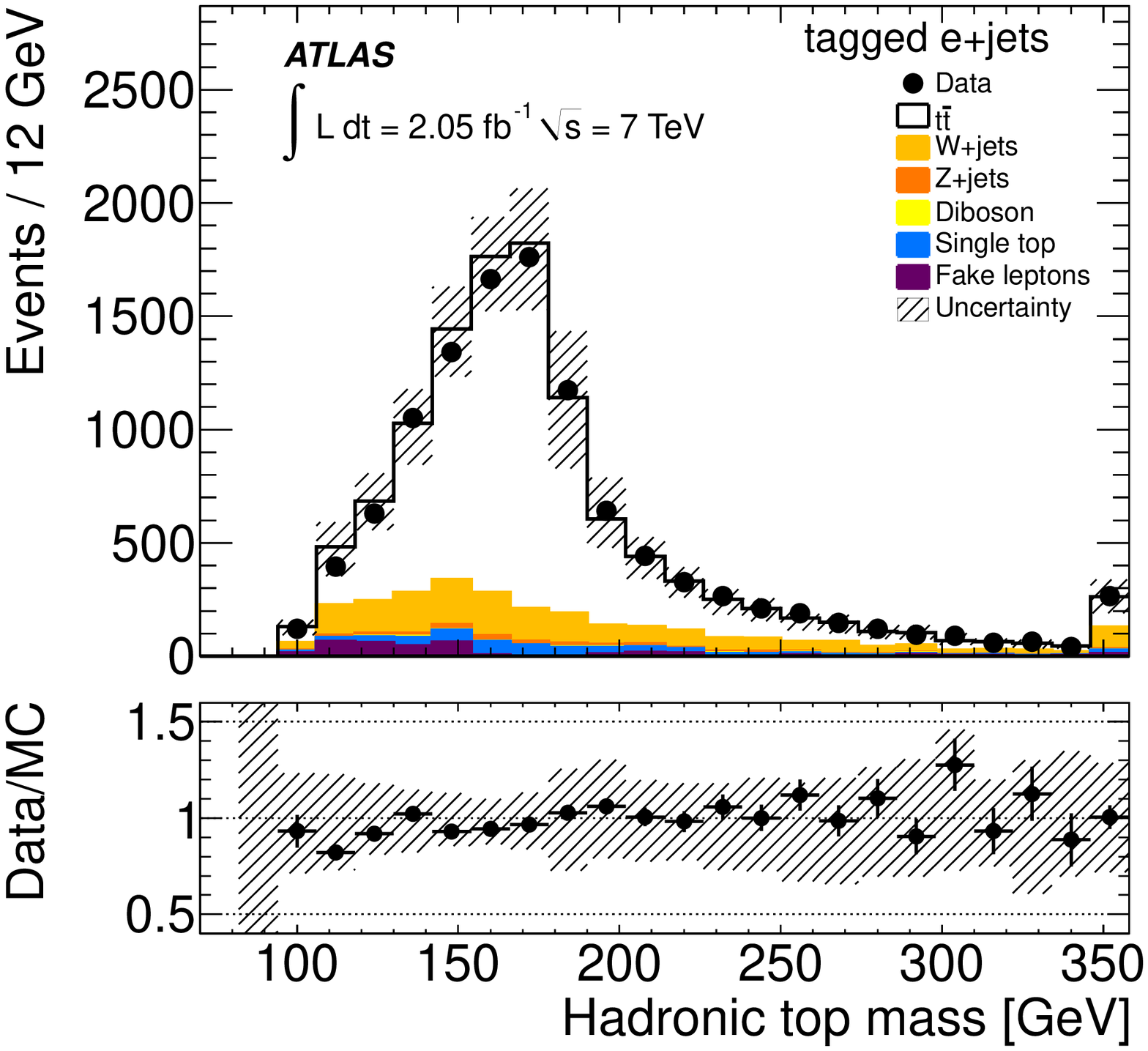}
      }
      \subfigure[]{
	\includegraphics[scale=0.37]{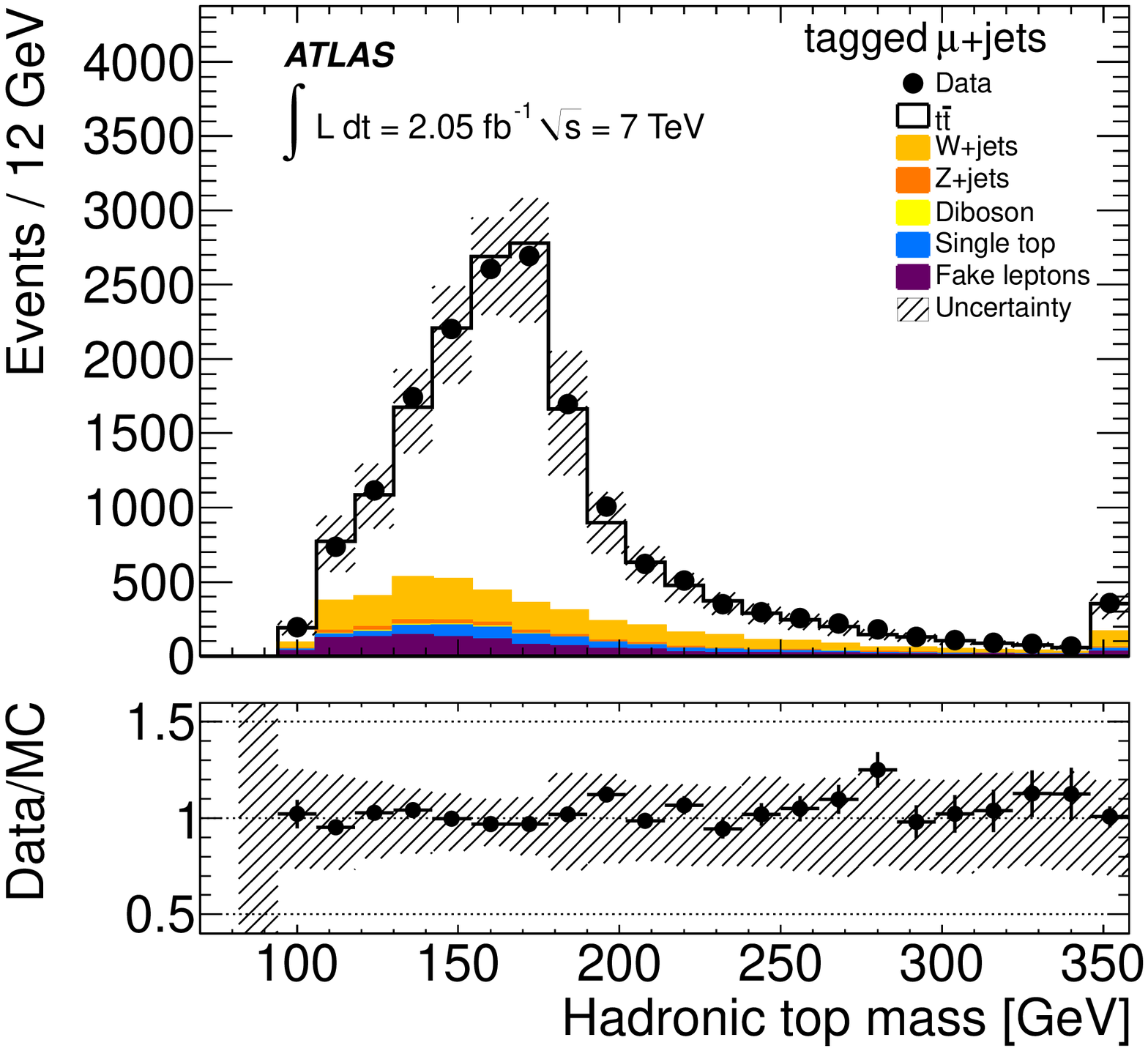}
      }

      \caption{Distributions of (a-b) jet multiplicity, (c-d) negative logarithm of the likelihood obtained from the
	kinematic fit described in the text and (e-f) invariant mass of the three reconstructed objects assigned to the
  hadronic top quark decay, obtained from the kinematic fit by relaxing
  the requirement on the value of the top quark mass (here named
 Hadronic top mass). In (c-d) the bin corresponding to the largest
	$-\ln(\textrm{likelihood})$ value includes events with
        $-\ln(\textrm{likelihood}) > 70$ and
	the associated	prediction.
In (e-f) the bin corresponding to the largest
Hadronic top mass value includes events with
      Hadronic top mass > 346 GeV and
	the associated	prediction.
 In (e-f) the top quark mass pole value is set to be the
    same in the Breit-Wigners describing the masses of the leptonic
    and hadronic top quarks, but it is not fixed to
    the value of 172.5 GeV. Data are compared to expectation from Monte Carlo
	simulation and data-driven expectation.  All selection
        criteria are applied, except for (a-b) for which only the
	likelihood requirement and the requirement on jet multiplicity
        are not applied and for (c-d)  for which only the likelihood requirement  is not applied.
 The band represents the 68\%
	confidence level interval of total uncertainty on the
	prediction. }
      \label{fig:lhnjetmass}
    \end{center}
  \end{minipage}
\end{figure*}

The likelihood is maximized by varying the energies of the partons, the
energy of the charged lepton, and the components of the
neutrino three-momentum. The maximization is performed over all
possible assignments of jets to partons, and the assignment with the
largest likelihood is used for all further studies. The
  distributions of the jet multiplicity are shown in Fig.~\ref
  {fig:lhnjetmass} (a-b) after all selection requirements, with the
  exception of the  requirements on the likelihood and  on the jet multiplicity.
The four-momenta of the top quarks are then obtained by summing the four
momenta of the decay products resulting from the kinematic fit.
The unconstrained $z$ component of the neutrino momentum is a free parameter
in the fit.

\begin{figure*}[htbp]
  \begin{minipage}{\textwidth}
    \begin{center}
      \subfigure[] {
	\includegraphics[scale=0.37]{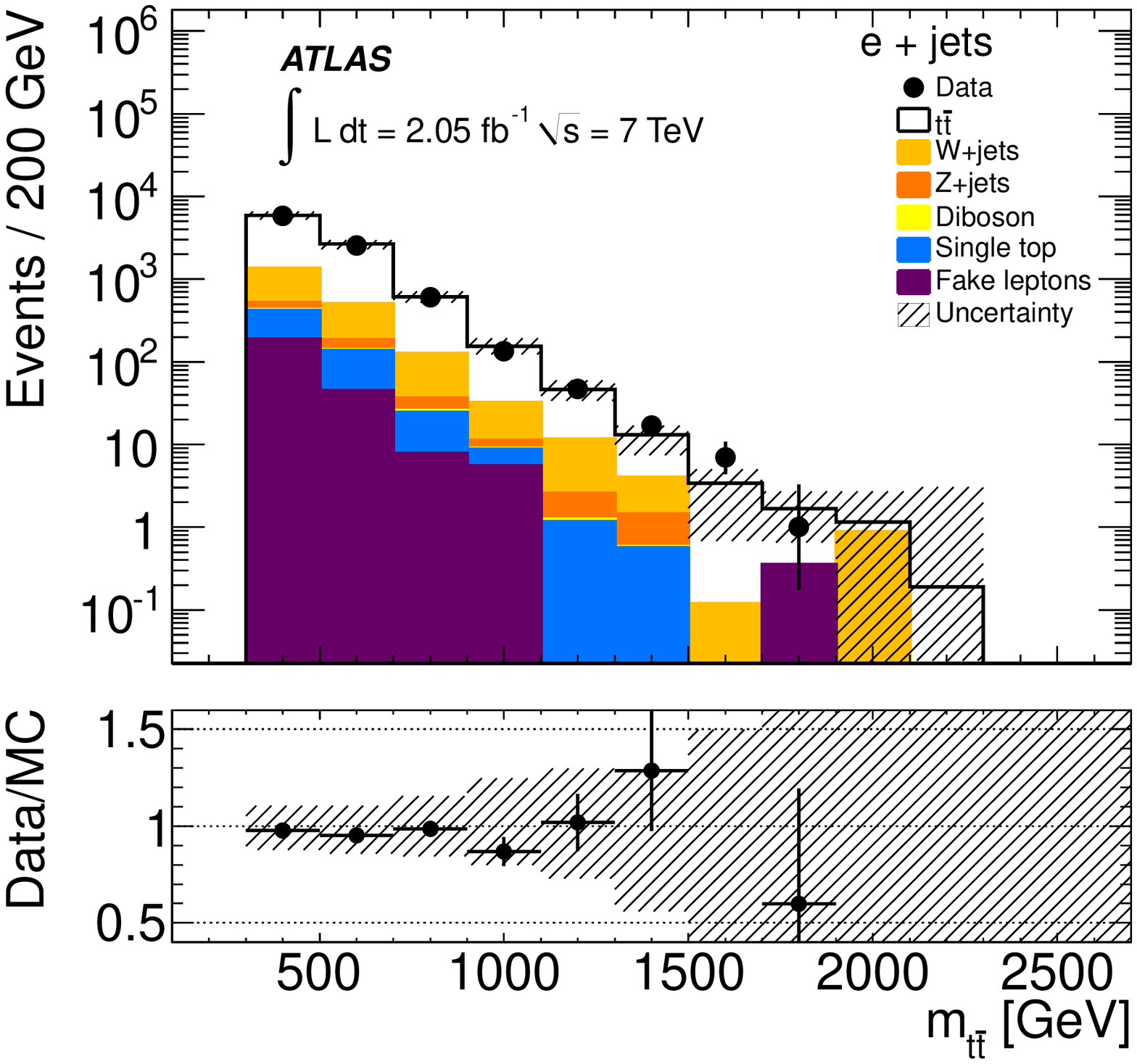}
      }
      \subfigure[]{
	\includegraphics[scale=0.37]{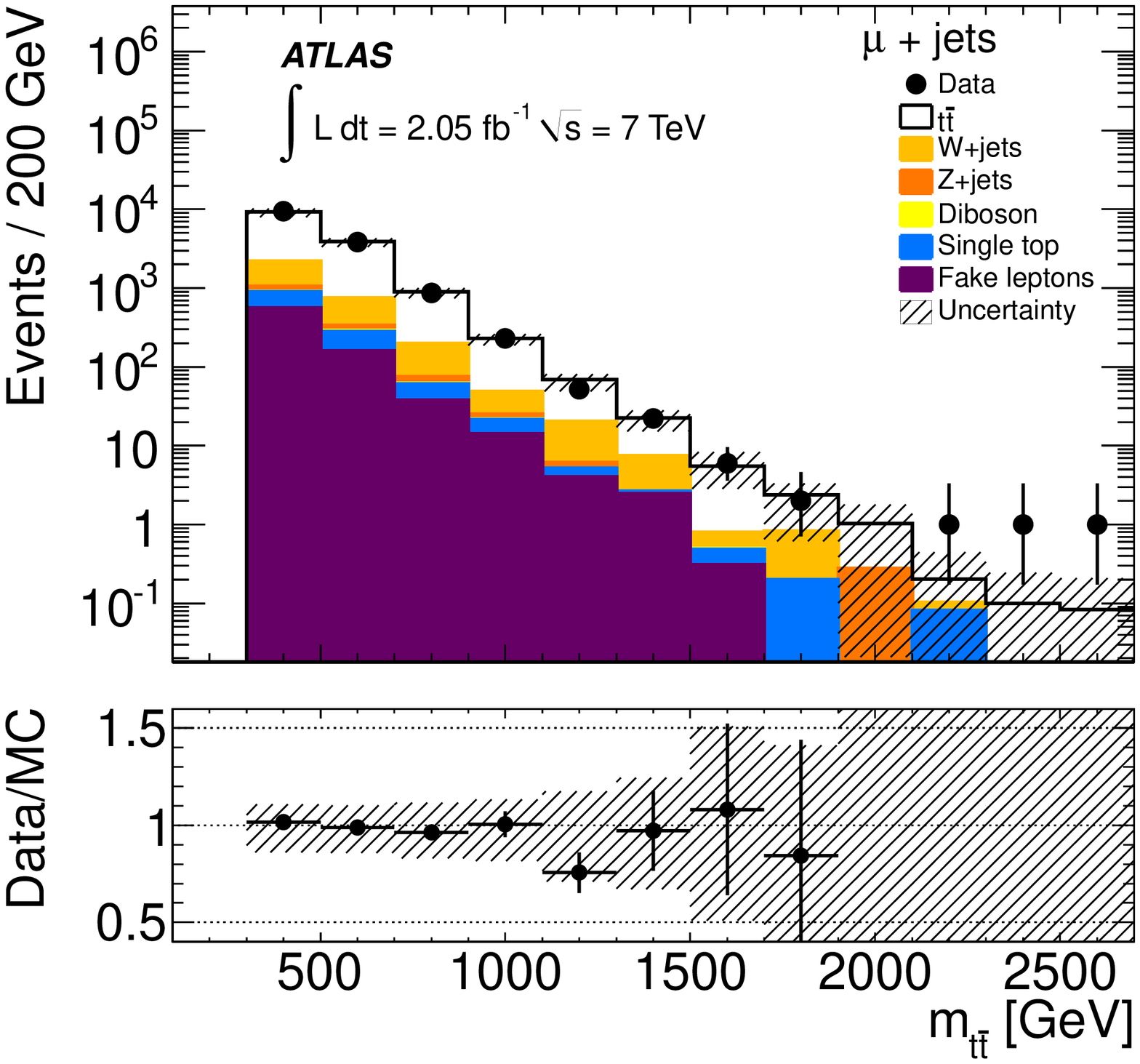}
      }
      \subfigure[]{
        \includegraphics[scale=0.37]{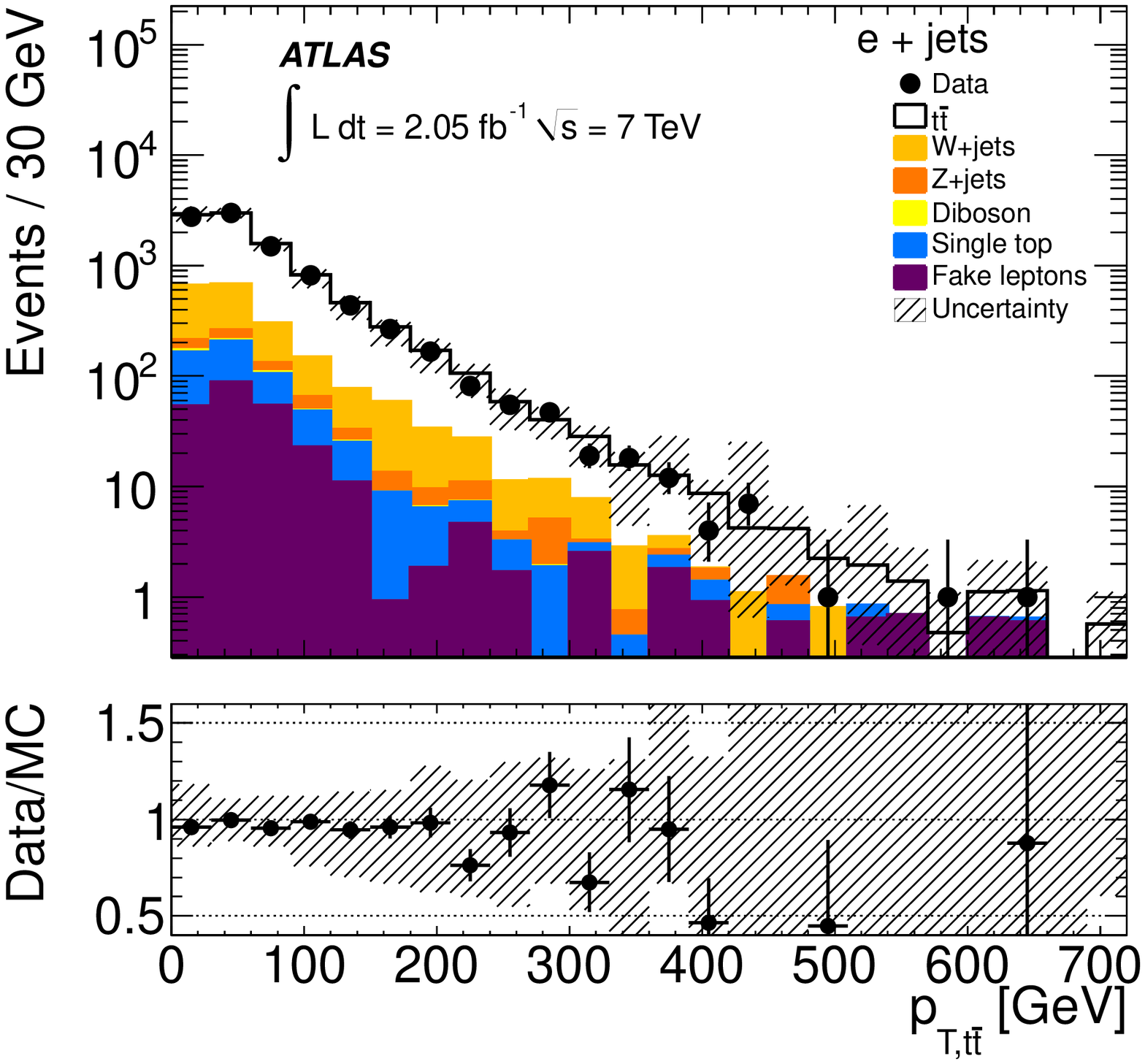}
      }
      \subfigure[]{
        \includegraphics[scale=0.37]{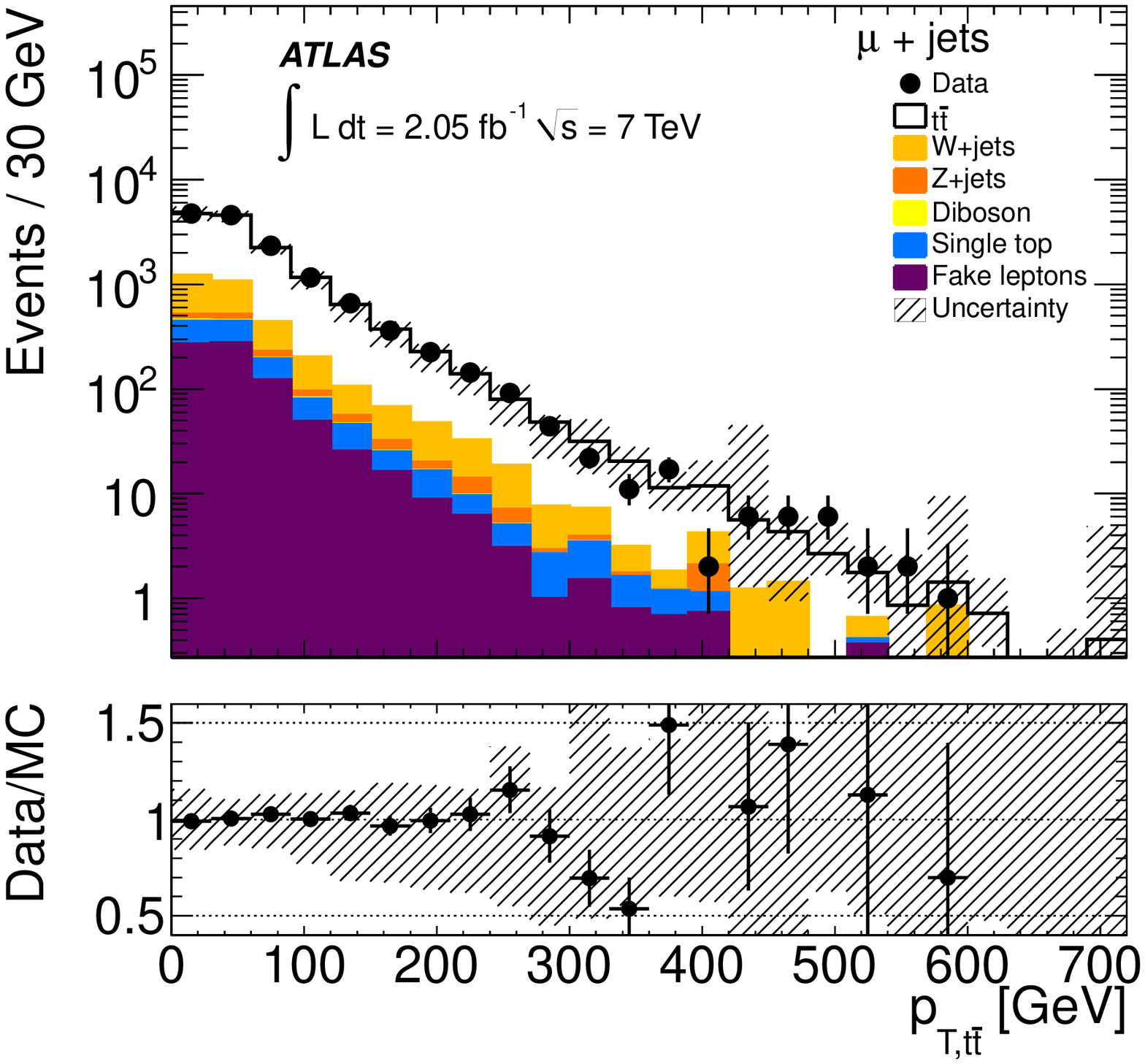}
      }
      \subfigure[]{
        \includegraphics[scale=0.37]{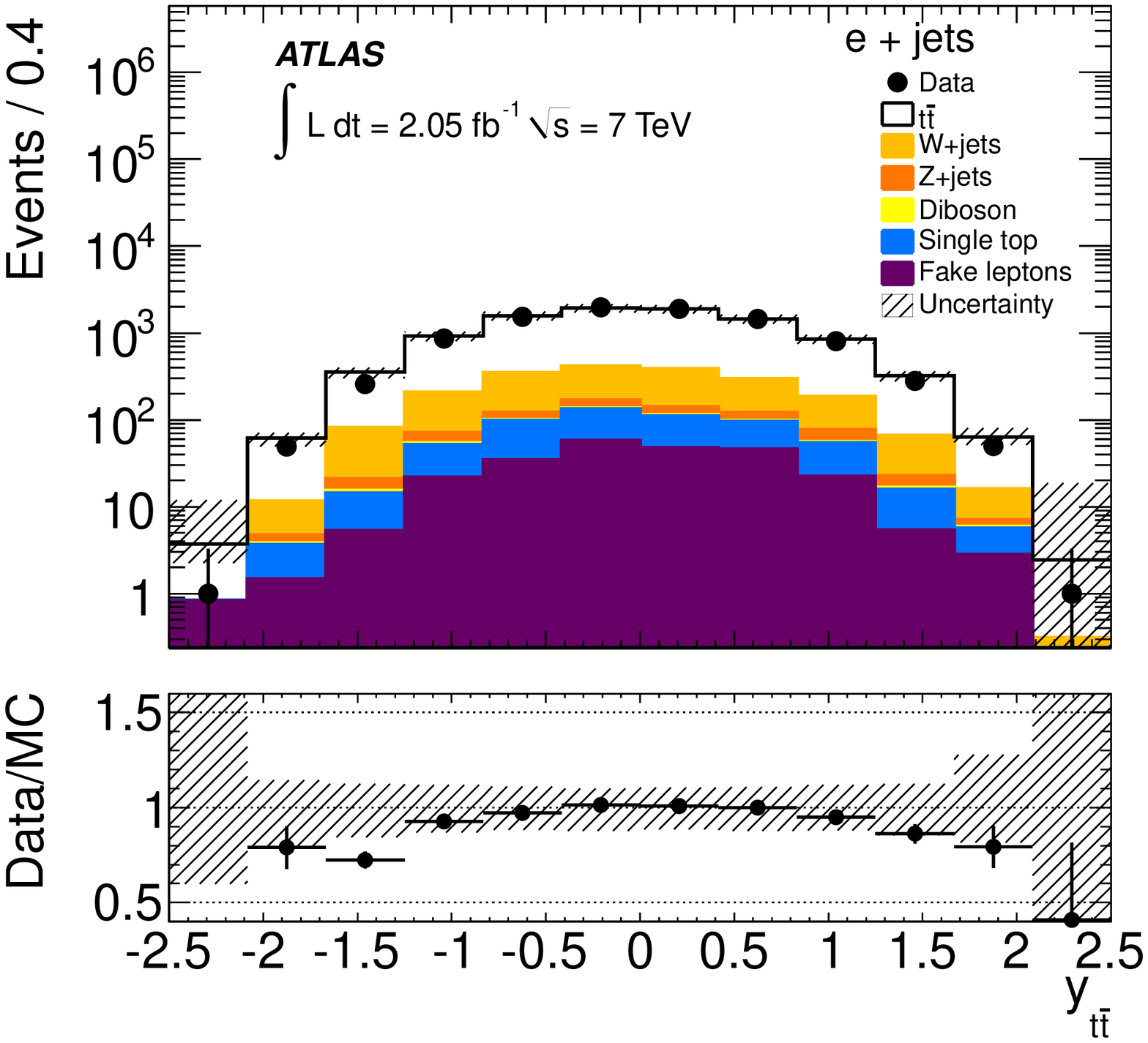}
      }
      \subfigure[]{
        \includegraphics[scale=0.37]{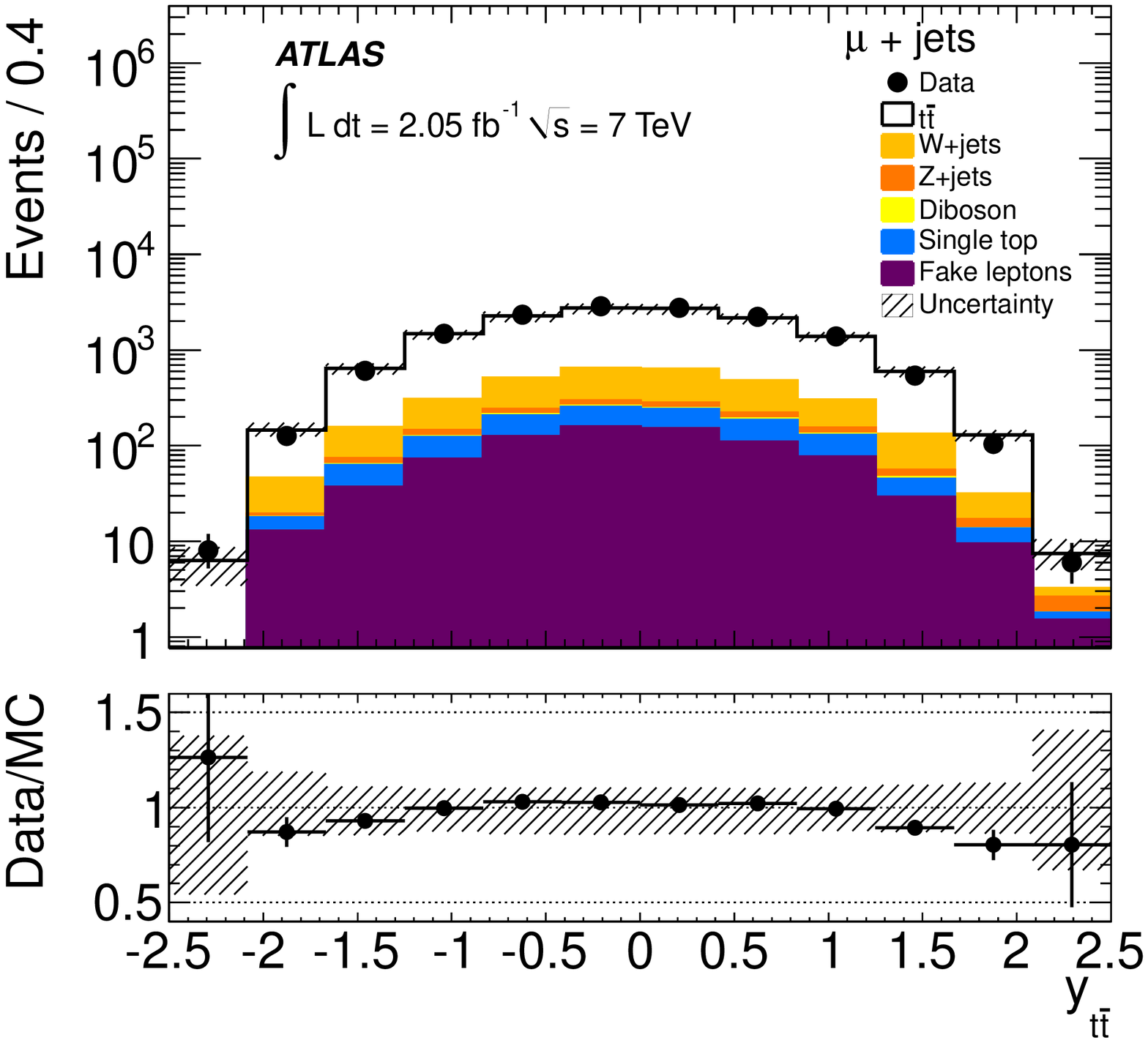}
      }
    \end{center}
  \end{minipage}
  \caption{
    Distributions of the  reconstructed (a-b) \ttbar\ mass,$\mttbar$,
    (c-d) the \ttbar\ transverse momentum,$p_{\mathrm{T},\ttbar}$,
    and (e-f) the \ttbar\ rapidity, $y_{\ttbar}$,
    before background subtraction and unfolding. 
    In (a-b) and (c-d) the bin corresponding to the largest $\mttbar$ (
    $p_{\mathrm{T},\ttbar}$)  value includes events with  $\mttbar$
    ($p_{\mathrm{T},\ttbar}$) larger than 2700 GeV (700 GeV).
    The largest reconstructed  $\mttbar$  in the \muplus\ channel is 2603 GeV.
        Data are compared to the
    expectation derived from simulation and data-driven estimates.
   All selection criteria are applied for the (a, c, e) \eplus and
    (b, d, f) \muplus channels.
    The uncertainty bands include all contributions given in
    Sect. \ref{sec:Systematics} except those from PDF and theory
    normalization.  } 
  \label{fig:m_p_rap_ttbar}
\end{figure*}

Simulation studies aimed at
enhancing the fraction of reconstructed \ttbar\ events that are
  consistent with the \ttbar\ decay assignment hypothesis are
used to determine a requirement on the likelihood of the kinematic fit. The
likelihood distribution for the events after selection, except for the
likelihood requirement $\ln(L) > -52$, is shown in
Fig.~\ref{fig:lhnjetmass} (c-d).
The likelihood optimally encapsulates all relevant information about the
data agreement with simulation. Figures~\ref{fig:lhnjetmass} (e-f)  show the distributions of the
 invariant mass of the three reconstructed objects assigned to the
  hadronic top quark decay, obtained from the kinematic fit by relaxing
  the requirement on the value of the top quark mass, after all
  selection requirements. In these distributions the top quark mass pole value is set to be the
    same in the Breit-Wigners describing the masses of the leptonic
    and hadronic top quarks, but it is not fixed to
    the value of 172.5 GeV. Further studies on the performance of the
    kinematic fit can be found in Ref.~\cite{ATLAS:2012aj}.
Distributions of the reconstructed invariant mass, transverse momentum and
rapidity of the reconstructed top-antitop pair, after all selection
requirements, are shown in Fig.~\ref{fig:m_p_rap_ttbar}.

The numbers of expected and observed data events in each
channels after pre-tag, tagged and full event selection are listed in
Table~\ref{evtnumbers}.  The data  agrees with the expectation within
the systematic uncertainties.

\begin{table*}[!htbp]
  \caption{Numbers of predicted and observed events.
    The selection is shown after applying pre-tag, tagged, and the full
    selection criteria including the likelihood requirement.
    The quoted uncertainties include all uncertainties given in Sect.
    \ref{sec:Systematics} except those from PDF and theory normalization.
    The numbers correspond to an integrated luminosity of \lumitot\ in
    both \eplus and \muplus samples.  }
  \label{evtnumbers}
  \begin{center}
    {
      \begin{tabular}{lr@{ $\pm$ }lr@{ $\pm$ }lr@{ $\pm$ }lr@{ $\pm$ }lr@{ $\pm$ }lr@{ $\pm$ }l}
	\hline
	Channel & \multicolumn{2}{c}{\muplus  pre-tag} & \multicolumn{2}{c}{\muplus tagged} &
	\multicolumn{2}{c}{\muplus L-req} &
	\multicolumn{2}{c}{\eplus pre-tag} & \multicolumn{2}{c}{\eplus tagged} &
	\multicolumn{2}{c}{\eplus L-req} \\
	\hline \hline
	$\ttbar$    & 15800 & 1300   & 13900  &  1100  & 11100  & 700  & 10700  & 900 & 9400 & 800 & 7400  & 500 \\
	\hline
	\wjets\   & 19000 & 5000   & 3000   &  1200    &  1700  & 700  & 13000  & 3300 & 2200 & 900 & 1300  & 500 \\
	\hline
	Single top   & 950  & 70       & 760    &  80   &  490 &  50   &    660   & 50 & 530  & 50 & 338   & 32\\
	\hline
	\zjets    & 2200  & 200   & 309    &  34        &  192  &  20  & 1750   & 330 & 240 & 50 & 154  &  26 \\
	\hline
	Diboson     & 298   & 28       & 53   &    7        &     34  &  4   &  181   & 19 & 32 & 5 & 21  &    3 \\
	\hline	      
	Fake-leptons  & 3400  & 1700   & 1100   &  1100   &  800  &  800&  2000   & 1000 & 400 & 400 & 250  &   250 \\
	\hline
	Signal+bkg   & 42000 & 6000   & 19200 &  2600   & 14400 & 1700  & 28000  & 4000 & 12800 & 1700 & 9500 &  1100 \\
	\hline
	Observed & \multicolumn{2}{c}{42327} &
	\multicolumn{2}{c}{19254} &
	\multicolumn{2}{c}{14416} & \multicolumn{2}{c}{26488} &
	\multicolumn{2}{c}{12457} &    \multicolumn{2}{c}{9187}
	\\ \hline \hline
      \end{tabular}
    }    
  \end{center}

\end{table*}

\section{Systematic uncertainties} 
\label{sec:Systematics}

For each systematic effect the analysis is re-run with the variation
corresponding to the one standard deviation change in each bin.
The varied distributions are obtained for the upward and downward shift
for each effect, and for each channel separately.
If the direction of the variation is not defined (as in the case of
the estimate resulting from the difference of two models), the
estimated variation is assumed to be the same size in the
upward and the downward direction and is symmetrized.
The baseline distribution and the shifted distributions
are the input to the pseudo-experiment calculation
(see Sect.~\ref{sec:results}) that performs unfolding,
efficiency correction, and enables combination of the \eplus and
\muplus channels. 

The sources  of systematic uncertainties are arranged in approximately independent groups.
They are further categorized into detector modelling,
and modelling of signal and background processes.
The estimation of the variations resulting from the systematic uncertainty
sources is the same as Ref.~\cite{CHARASY-012}.

\subsection{Detector modelling}
\label{sec:SystematicsDet}

Muon and electron trigger, reconstruction and
selection efficiencies are measured in data using $Z$ and $W$ decays
and incorporated into the simulation using weighted events. Each simulated event
is weighted with the appropriate ratio (scale factor) of the measured efficiency to
the simulated one.
The uncertainties on the scale factors are estimated 
by varying the lepton and signal selections and background uncertainties.
For lepton triggers the systematic uncertainties are about $1\%$.
The same procedure is used for lepton momentum scale
and resolution scale factors resulting in uncertainties of 1--1.5\%.
The corresponding scale factor uncertainties for electron (muon)
reconstruction and
identification efficiency are 1\% (0.5\%) and 3\% (0.5\%) respectively.

Information collected from collision data, test-beam data, and simulation is
used to determine the jet energy scale; its uncertainty ranges from
2.5\% to 8\%, varying with jet \pt\ and $\eta$~\cite{Aad:2011he}.
The uncertainties include
flavour composition of the sample and mis-measurements due to nearby
jets. Pile-up gives an additional uncertainty of 2.5\%
(5\%) in the central (forward) region. An extra uncertainty of up to
2.5\% is added to account for the fragmentation of $b$-quarks.
The jet energy resolution and reconstruction efficiency are measured in
data using the same methods as in Refs.~\cite{Aad:2011he,ATLAS-CONF-2010-054}.
Jet energy resolution uncertainties range from 9--17\%  for jet
\pt\ $\simeq$ 30 \gev\ to  about 5--9\%  for  jet \pt\ $>$
180 \gev\ depending on jet \eta. The jet reconstruction
efficiency uncertainty is 1--2\%. The uncertainties from the
energy scale and resolution corrections on leptons and jets are
propagated to the uncertainties on missing transverse momentum.
Uncertainties on \met{} also include contributions arising from
calorimeter cells not associated to jets and from soft jets (those in the
range 7~\gev\ $<$ \pt\ $<$ 20~\gev). The $b$-tagging efficiency scale factors
have
uncertainties between 6\% to 15\%, and mis-tag rate scale factor uncertainties
range from 10\% to 21\%. The scale factors are derived from data and
parameterized as a function of jet \pt.

A small region of the liquid argon calorimeter could not be read out
in a subset of the data corresponding to 42\% of the total dataset.
Corresponding data and simulated events where a jet with $\pt>20$ \gev\ is
close to the failing region are rejected. This requirement rejects about
6\% of the events.
A systematic uncertainty is derived from variations of the
\pt-threshold of the jets by  20\% resulting from studies of
  the response of jets close to the failing region, using dijet \pt\ balance in data.

The uncertainty on the measured luminosity is
3.7\%~\cite{Aad:2011dr,ATLAS-CONF-2011-011}.  

\subsection{Signal and background modelling}
\label{sec:SystematicsMod}

Sources of systematic uncertainty for the signal
are the choice of generator, parton shower model, hadronization
and underlying event model, the choice
of PDF, and the tuning of initial- and final-state
radiation. Predictions from the  \mcnlo{} and
\powheg~\cite{Nason:2004rx,Frixione:2007vw} generators are
compared to determine the generator dependence. The parton showering is assessed by comparing
\powheg\ samples interfaced to \herwig\ and \pythia,
respectively. The a\-mount of initial- and final-state radiation is
varied by modifying
parameters in \acermc~\cite{AcerMC35} interfaced to \pythia.
The parameters are varied in a range comparable to
those used in the Perugia Soft/Hard tune
variations~\cite{Skands:2010ak}. 
The present initial-state radiation variations are to be
  considered generous: the spread of the resulting theoretical
  predictions for jet activity in \ttbar\ events is often wider than
  the experimental uncertainties in precision measurements performed
  by ATLAS in LHC $pp$ collisions at \rts\ = 7 \tev~\cite{ATLAS:2012al}.
The impact of the PDF uncertainties is studied using the procedure
described in Ref.~\cite{Botje:2011sn,Nadolsky:2008zw,MSTW2008nlo,NNPDF2010}.

Background processes are
either estimated by simulation or using auxiliary measurements,
see Sect.~\ref{sec:bkg}. The uncertainty on the fake-lepton background
is estimated by varying the requirements on the low-\mtw\ and
low-\met\ control regions,
taking into account the statistical uncertainty and background corrections.
The total uncertainty is estimated to be 100\%.
The normalization of \wjets\ is
derived from auxiliary measurements using the asymmetric production
of positively and negatively charged $W$ bosons in \wjets\
events. The total uncertainties are estimated to be 21\% and 23\% in the four-jet bin,
for the electron and muon channels respectively.
These uncertainties are estimated by evaluating the effect on both
$r_{\rm MC}$ and $k_{2\to\ge4}$ 
from the jet energy scale uncertainty and different PDF and generator choices.
Systematic uncertainties on
the shape of \wjets\ distributions are assigned based on differences
in simulated events generated with different factorization and
parton matching scales.
Scaling factors correcting the fraction of heavy flavour
contributions in simulated \wjets\ samples are derived from
auxiliary measurements (see Sect.~\ref{sec:ChAsymm}). The systematic
uncertainties are found by changing the normalizations of the non-$W$ processes
within their uncertainties when computing $W^{\rm data}_{i,
  \textrm{pre-tag}}$ and $ W^{\rm data}_{i, \rm tagged}$,
as well as taking into account the impact of uncertainties in $b$-tagging
efficiencies.
The uncertainties are 47\% for \wbbjets\ and \wccjets\
contributions and 32\% for \wcjets\ contributions. 
In the \muplus channel the fractional
contributions of \wbbjets, \wccjets\ and \wcjets\ samples to the total \wjets\ prediction are estimated to be 9\%, 17\%
and 12\% (36\%, 25\% and 17\%)  respectively, before (after) the
$b$-tagging requirement.
In the \eplus channel the fractional
contributions of \wbbjets, \wccjets\ and \wcjets\ samples to the total \wjets\ prediction are estimated to be 9\%, 17\%
and 13\% (35\%, 25\% and 17\%)  respectively, before (after) the
$b$-tagging requirement.
The normalization of \zjets\ events is
estimated using Berends-Giele-scaling~\cite{Berends199132}. The
uncertainty in the normalization is 48\% in the four-jet bin and
increases with the jet multiplicity. The uncertainties on the
normalization of the small background contributions from diboson and single top
production are estimated to be about 5\%~\cite{MCFM_DIBOS,MSTW2008a,MSTW2008b} and
10\%~\cite{KIDO-011,KIDO-010a,KIDO-010b}, respectively.

The statistical uncertainty on the Monte Carlo prediction due to
limited Monte Carlo sample size is included as a systematic uncertainty
in each bin for each process.

\section{Cross-section unfolding}
\label{sec:unfolding}

\subsection{Unfolding procedure}

The underlying binned true differential cross-section distributions
($\sigma_j$)
are obtained from the reconstructed events using an unfolding technique
that corrects for detector effects.
The unfolding starts from the reconstructed event distribution ($N_i$),
where the backgrounds ($B_i$) have been subtracted. The unfolding uses
a response matrix ($R_{ij}$), see Eq. (\ref{eq:fold}), derived from
simulated \ttbar\ events, which maps
the binned generated events to the binned reconstructed events.
The kinematic properties of the generated $t$ and \tbar\ partons in simulated
\ttbar\ events define the ``true'' properties of the \ttbar\ events.
 
\begin{figure*}[htbp]
  \begin{minipage}{\textwidth}
    \begin{center}
      \subfigure[]{
	\includegraphics[scale=0.37]{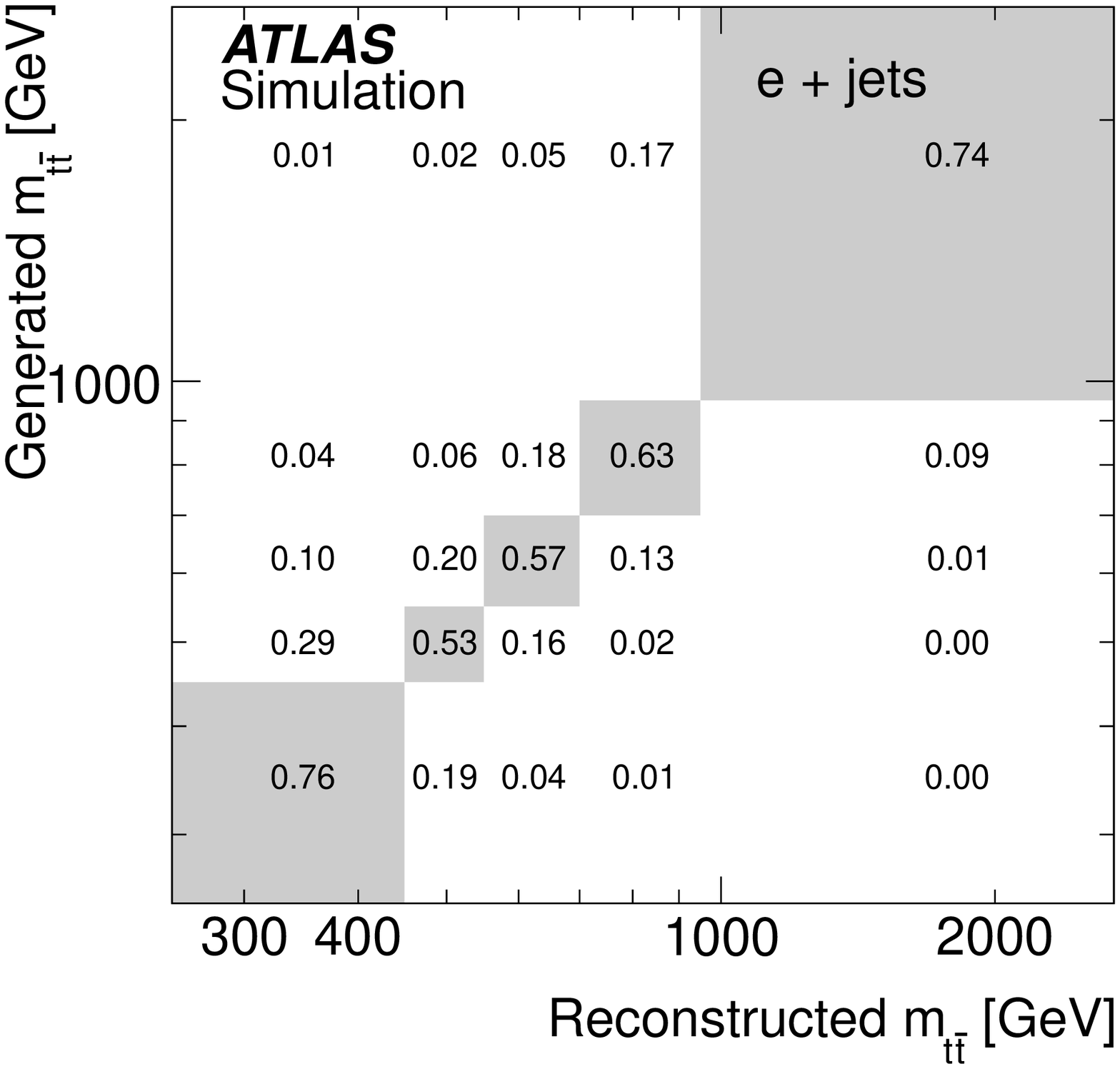}
      }
      \subfigure[]{
	\includegraphics[scale=0.37]{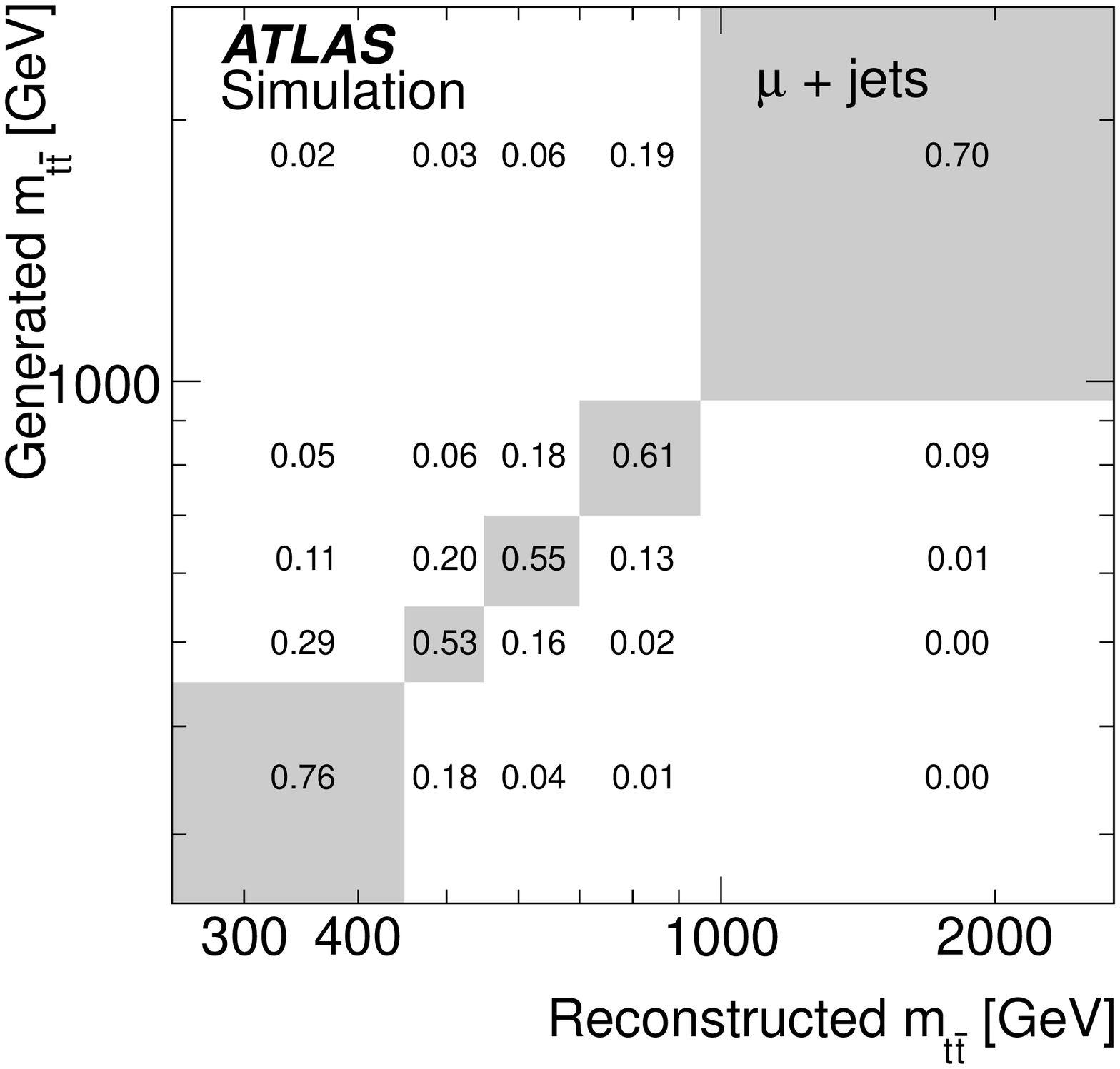}
      }
      \subfigure[]{
	\includegraphics[scale=0.37]{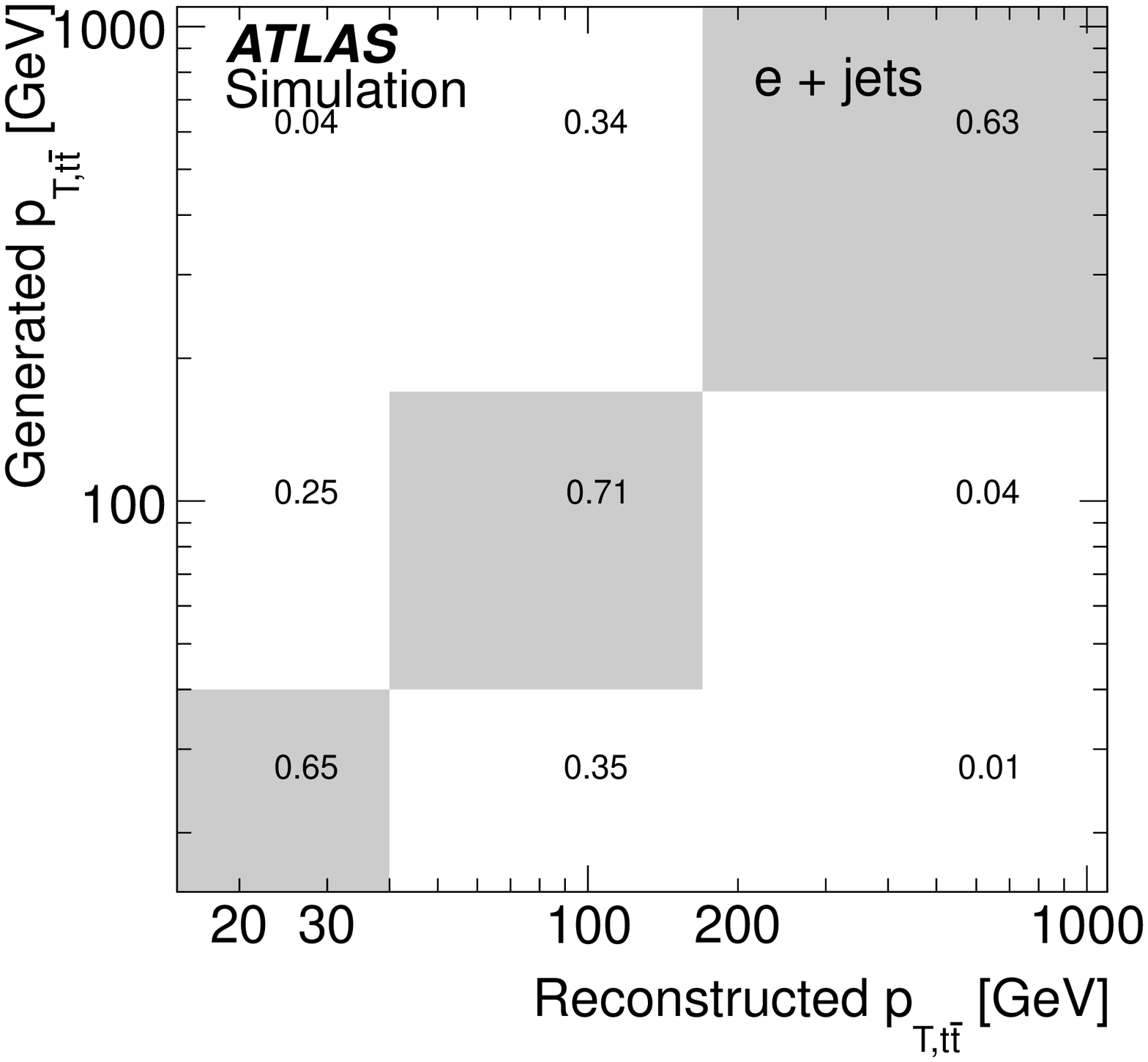}
      }
      \subfigure[]{
	\includegraphics[scale=0.37]{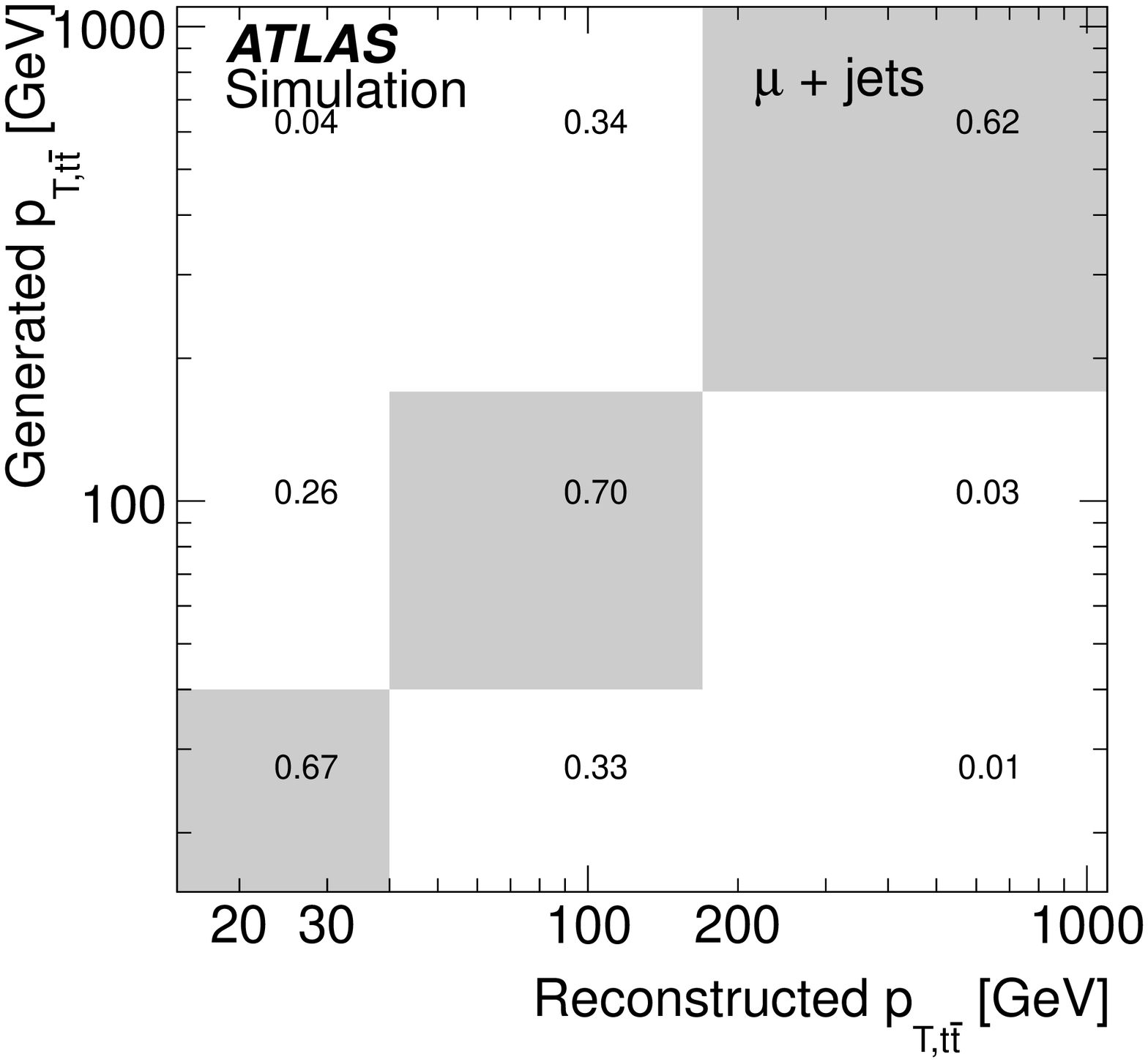}
      }
      \subfigure[]{
	\includegraphics[scale=0.37]{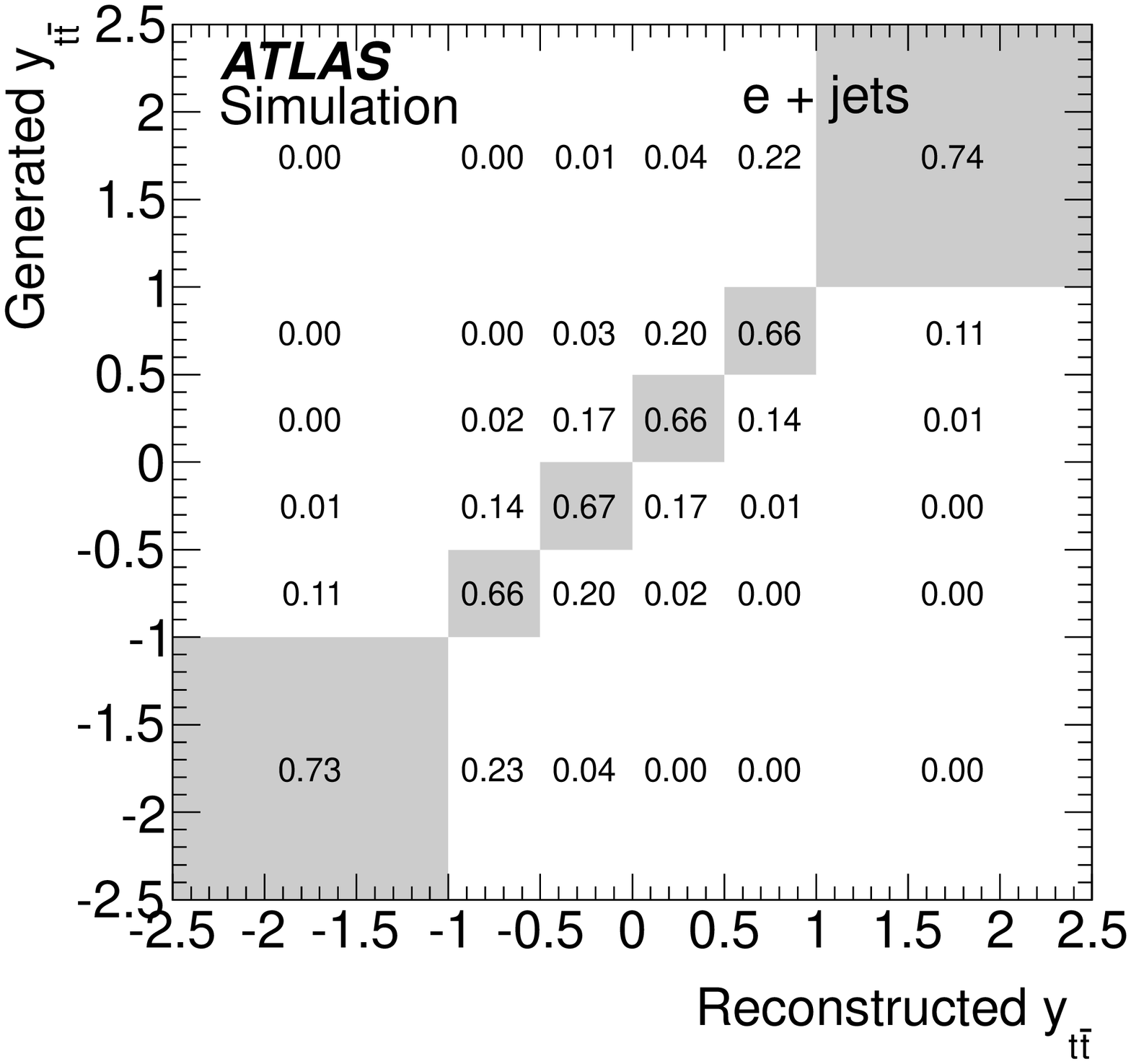}
      }
      \subfigure[]{
	\includegraphics[scale=0.37]{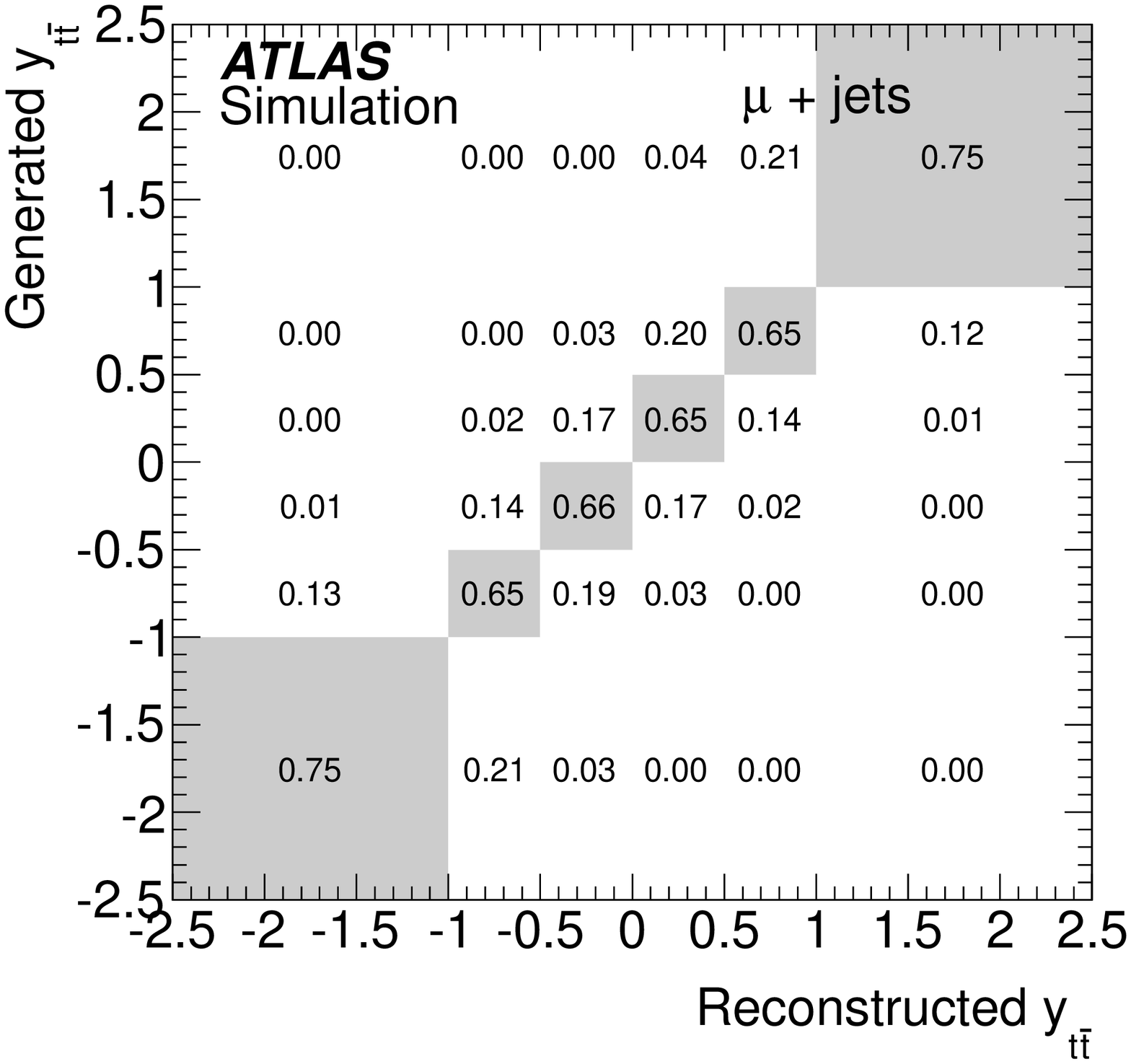}
      }
    \end{center}
  \end{minipage}
  \caption{Migration matrices for (a-b) $\mttbar$, (c-d) $p_{\mathrm{T},\ttbar}$,
   and (e-f) $y_{\ttbar}$ estimated from  simulated \ttbar\ events
    passing all (left)\eplus and (right) \muplus selection criteria. The unit
    of the matrix elements is the probability  for an event generated
    at a given value to be reconstructed at another value.
  }
  \label{fig:mig}
\end{figure*}

In its simplest form the unfolding equation can be written as
\begin{equation} \label{eq:fold}
  N_i = \sum_{j} R_{ij} \sigma_j \mathcal{L} + B_i = \sum_{j} M_{ij} A_j
  \sigma_j \mathcal{L}
  + B_i,
\end{equation}
where $\mathcal{L}$ is the integrated luminosity, $M_{ij}$ is the bin
migration matrix (see Fig.~\ref{fig:mig}), and  $A_j$ is the
acceptance for inclusive \ttbar\ events.  The leptonic branching
fractions are set according to  Ref.~\cite{PDG2010}.

\begin{figure*}[htbp]
  \begin{minipage}{\textwidth}
    \begin{center}
      \subfigure[] {
	\includegraphics[scale=0.37]{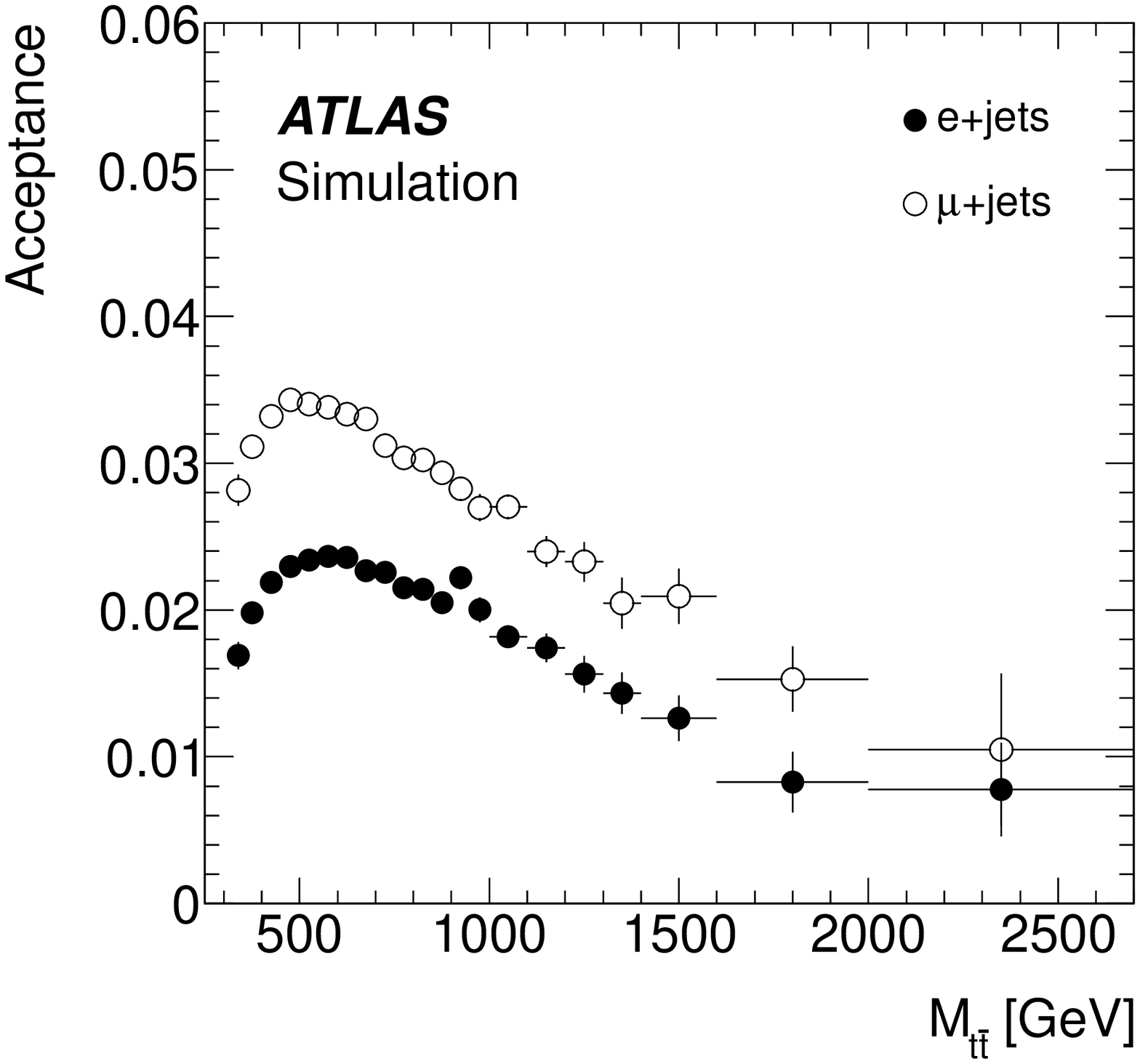}
      }
      \subfigure[]{
	\includegraphics[scale=0.37]{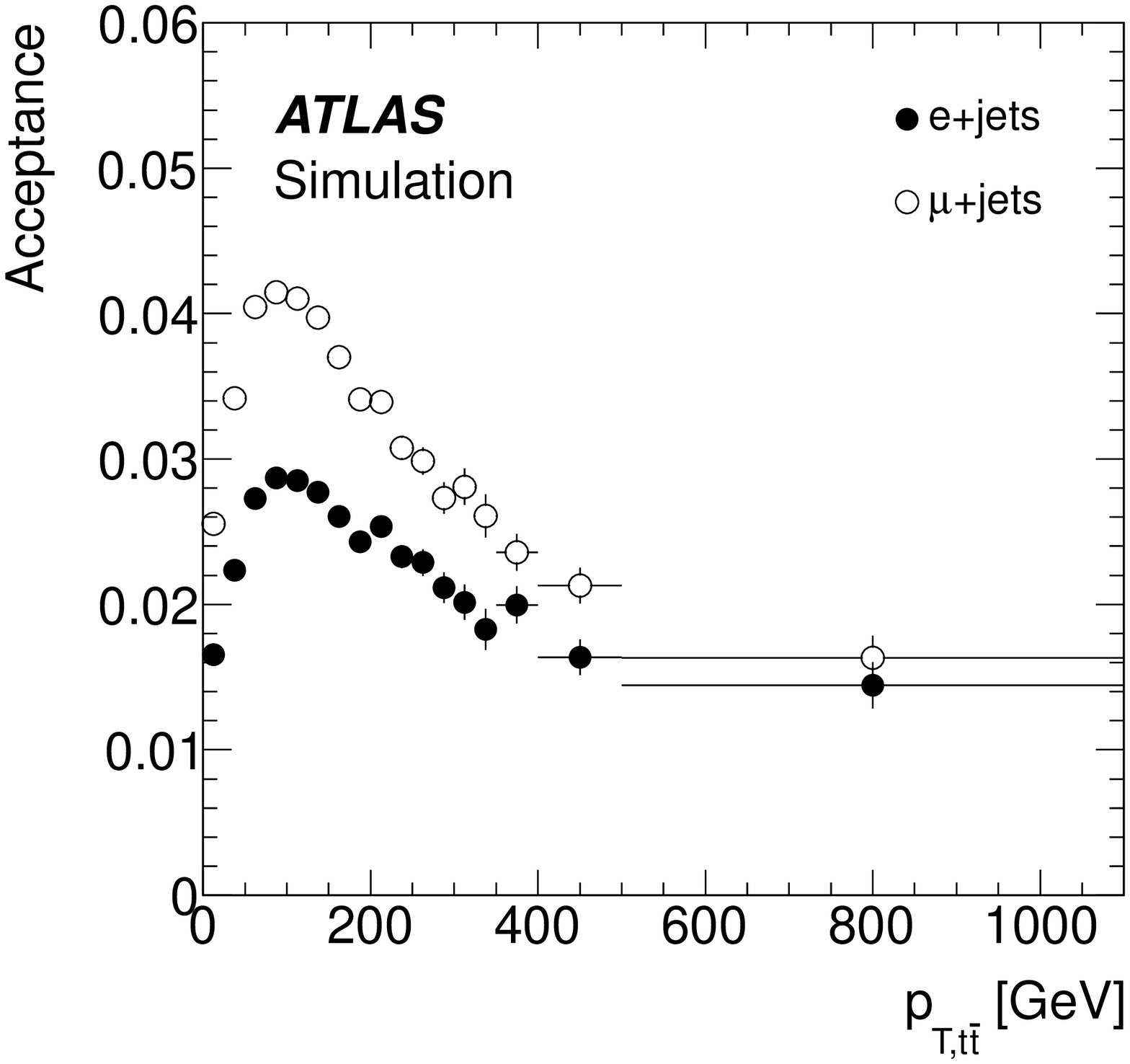}
      }
      \subfigure[]{
	\includegraphics[scale=0.37]{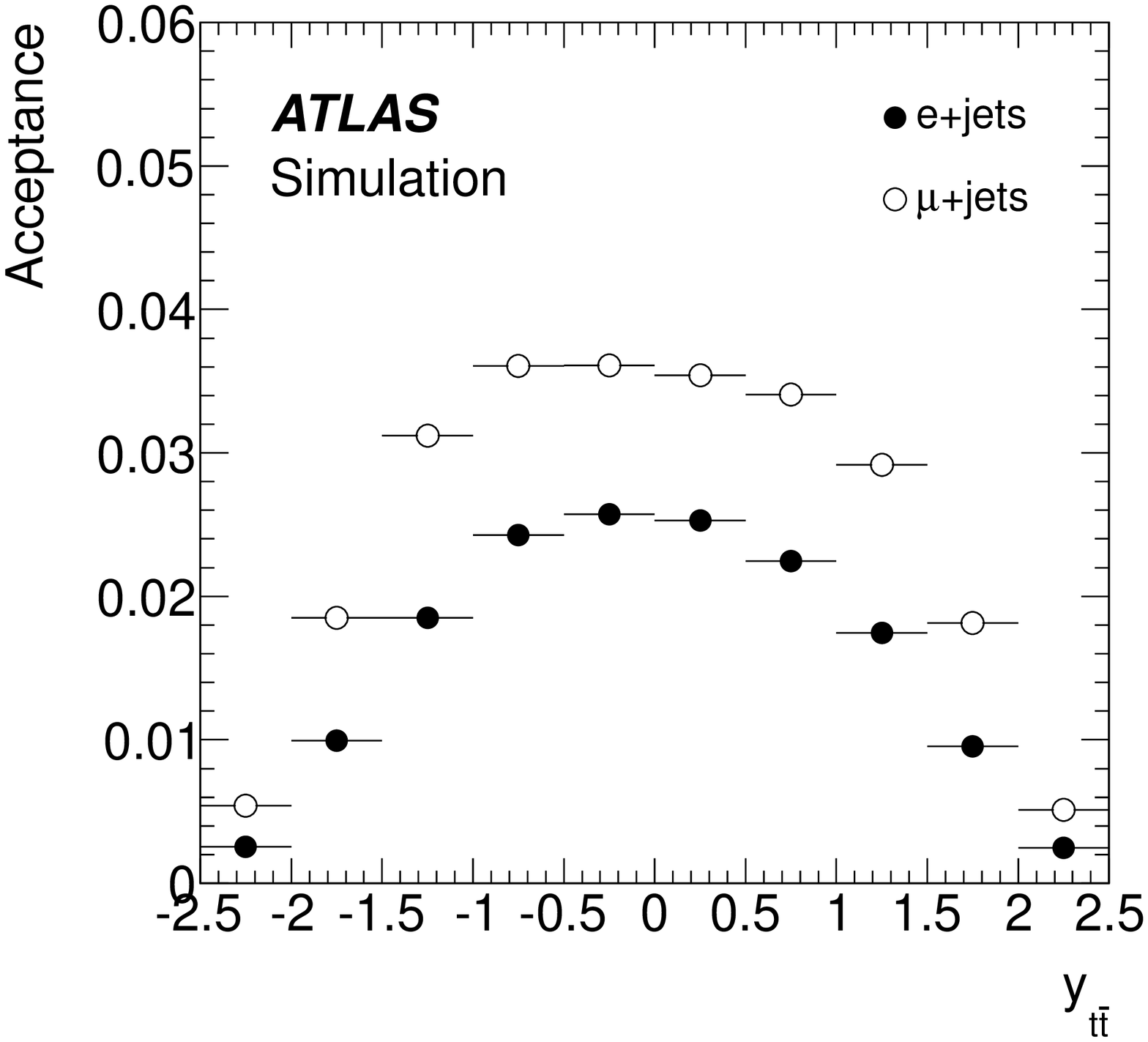}
      }
    \end{center}
  \end{minipage}
  \caption{
    Acceptance as a function of (a) \ttbar\ mass,
    $\mttbar$,
    (b) the \ttbar\ transverse momentum, $p_{\mathrm{T},\ttbar}$,
    and (c) the \ttbar\ rapidity, $y_{\ttbar}$.
    The acceptance is defined according to Eq.~\ref{eq:fold}  for inclusive \ttbar\ events after all
    selection requirements. The leptonic branching fractions are set
    according to  Ref.~\cite{PDG2010}.
The error bars show only the uncertainty due to
    limited Monte Carlo sample size.}
  \label{fig:acc_fine}
\end{figure*}

The estimated acceptances for simulated \ttbar\ events as a function of
\mttbar, $p_{\mathrm{T},\ttbar} $ and $ y_{\ttbar} $ are reported in
Table~\ref{tab:acceptanceMttbar}. The overall acceptances
before the requirement on the likelihood value are comparable to previous
measurements~\cite{Aad:2010ey}. The additional requirement on the likelihood
value is expected to retain a large fraction of the previously selected
\ttbar\ events (see Table~\ref{evtnumbers}).
A finely binned illustration of the acceptances is shown
in Fig.~\ref{fig:acc_fine}. The reduction in acceptance associated with
high \mttbar\ and $p_{\mathrm{T},\ttbar} $ values is
predominantly due to the presence of increasingly non-isolated leptons
coupled to lower jet multiplicity as \ttbar\ decay products are
  forced in a closer space region by the boost at large top quark \pt. In the case of high $|y_{\ttbar}|$ it
is mainly due to jets falling outside of the required pseudorapidity
range (see Sect.~\ref{evselection}).

\begin{table*}[thbp] 
  \caption{Table of acceptances for
    \mttbar, $p_{\mathrm{T},\ttbar}$ and $y_{\ttbar}$.
    The acceptance is defined according to Eq.~\ref{eq:fold}  for inclusive \ttbar\ events after all
    selection requirements. The leptonic branching fractions are set according to  Ref.~\cite{PDG2010}.
In the case of $y_{\ttbar}$, acceptances
      in positive and negative symmetric bins are consistent within uncertainties.}
  \label{tab:acceptanceMttbar}

  \begin{center}
    \begin{minipage}[t]{0.3\textwidth}
      \vspace{0pt}
      \begin{tabular}{c c c}\hline
	$\mttbar$ [\gev] & \multicolumn{2}{c}{ Acceptance [\%]} \\  \hline
	& \eplus & \muplus \\ \hline\hline
	250 -- 450 & 2.1 & 3.2 \\ 
	450 -- 550 & 2.3 & 3.4 \\ 
	550 -- 700 & 2.4 & 3.4 \\ 
	700 -- 950 & 2.2 & 3.1 \\ 
	950 -- 2700 & 1.8 & 2.5 \\ \hline\hline
      \end{tabular}
    \end{minipage}    
    \begin{minipage}[t]{0.3\textwidth}
      \vspace{0pt}
      \begin{tabular}{c c c}\hline
	$p_{\mathrm{T},\ttbar} $ [\gev] &\multicolumn{2}{c}{Acceptance [\%]}\\\hline
	& \eplus & \muplus \\ \hline\hline
	0 -- 40 & 1.8 & 2.8 \\ 
	40 -- 170 & 2.7 & 4.0 \\ 
	170 -- 1100 & 2.3 & 3.1 \\ \hline\hline
      \end{tabular}
    \end{minipage}
    \begin{minipage}[t]{0.3\textwidth}
      \vspace{0pt}
      \begin{tabular}{c c c}\hline
	$ y_{\ttbar} $&  \multicolumn{2}{c}{Acceptance [\%]}    \\  \hline 
	& \eplus & \muplus \\ \hline\hline
	-2.5 -- -1.0 & 1.5 & 2.6 \\ 
	-1.0 -- -0.5 & 2.4 & 3.6 \\ 
	-0.5 -- 0.0 & 2.6 & 3.6 \\ 
	0.0 -- 0.5 & 2.5 & 3.6 \\ 
	0.5 -- 1.0 & 2.3 & 3.4 \\ 
	1.0 -- 2.5 & 1.5 & 2.5 \\ \hline\hline
      \end{tabular}
    \end{minipage}
    
 \end{center}
\end{table*}

The cross-section $\sigma_{j}$ is then extracted by solving
Eq.~(\ref{eq:fold})
\begin{equation} \label{eq:unfold}
  \sigma_j = \frac{\sum_{i} (M^{-1})_{ji} (N_i - B_i)}{A_j \mathcal{L}}.
\end{equation}
The bin size is optimized using pseudo-experiments drawn from
simulated events including systematic uncertainties. The adopted
optimization strategy is to choose as small a bin size as possible
without substantially increasing the total uncertainty after unfolding.
This effectively means keeping about
68\% of the events on the diagonal of the migration matrix,
and requiring that the condition number\footnote{The condition number
  $k$ is defined as $k=||M|| \cdot ||M^{-1}||$, and is a measure
  of how much the matrix inversion increases the size of the
  uncertainties in the error propagation.
}
of the migration matrix is $O(1)$.  The finely binned distributions
before unfolding reported in Fig.~\ref{fig:m_p_rap_ttbar} show good
agreement between reconstructed data and the MC and data-driven predictions.

To evaluate the performance of the unfolding procedure, and
to estimate the systematic uncertainties, Eq.~(\ref{eq:unfold})
has been extended to the following form to allow
detailed studies using pseudo-experiments
\begin{equation} \label{eq:psunfold}
  \sigma_j(d_k) = \frac{\sum_{i} (M^{-1})_{ji}(d_k)[P(N_i) - B_i(d_k)]}
	{A_j(d_k)\mathcal{L}(d_k)},
\end{equation}
where $P(N_{i})$ is the Poisson distribution with mean $N_{i}$, and $d_k$
are continuous variables representing the systematic uncertainties, drawn
from a Gaussian distribution with zero mean and unit standard deviation.
A cross-section estimate $\sigma_{j}$  is
extracted for a given variable ($\mttbar$,
$p_{\mathrm{T},\ttbar}$, $y_{\ttbar}$) from each pseudo-experiment. The
distribution of $\sigma_{j}$  resulting
from the pseudo-experiments is
an estimator of the probability density of all possible outcomes of
the measurement.
Two thousand pseudo-experiments are used to extract the cross-section
values. The 68\% confidence interval provides the cross-section uncertainty.
The parametric dependence on $d_k$ in $(M^{-1})_{ij}$,
and other functions, is approximated using the linear term in the
Taylor expansion, treating positive and negative derivative
estimates separately.

A closure test is performed by unfolding simulated (folded) events
where $d_k = 0$. The
deviation of the unfolded cross-section from the known true cross-section
input, used for the detector simulation folding,
is consistent with zero within 1\% uncertainty.
The most important test of the unfolding is to test the ability to unfold a
distribution significantly different from the Monte Carlo expectation.
This is done by re-weighting simulated \ttbar\ events so that the number of
events in a single bin of true $\mttbar$ is doubled.
The observed linearity of the response to these ``delta-like'' pulses
is within 1\%. The same test was also performed using a regularized unfolding
technique based on Singular Value Decomposition~\cite{HOE-9601}.
The size of the ``delta-like'' pulses was then found to be substantially
reduced (at least by 30\%) after unfolding, even under the mildest regularization
conditions. Given the bias from this particular unfolding implementation  which does not
allow to reduce the regularization any further, all final results are
derived using the plain matrix inversion described above. The
increased statistical uncertainty of this unregularized result is tolerated 
given that the total uncertainty is dominated by systematic effects.

\subsection{Combination of channels}

The unfolded cross-sections from the two channels, \eplus and \muplus, are
combined using a weighted mean which includes the full
covariance matrix between the channels. Since the covariance matrix
is used in the weighting, the estimate is a best linear
unbiased estimator of the cross-section. The covariance matrix
is determined in simulated events using the same pseudo-experiment procedure
outlined in the previous section and derived from
Eq.~(\ref{eq:psunfold}).

\section{Results}
\label{sec:results}

To reduce systematic uncertainties only relative
cross-sections (differential cross-section normalized to the measured inclusive
cross-section) are reported.  The full procedure for the differential
measurement is also contracted down to one bin to perform the measurement of 
the inclusive cross-section by using Eq.~(\ref{eq:fold})
and Eq.~(\ref{eq:unfold}). In this case the measurement is reduced to a
standard ``cut-and-count'' technique (as used 
for  the first ATLAS \ttbar\ cross-section
measurement~\cite{Aad:2010ey}) and the response matrix is
reduced to the standard acceptance correction. 
The total inclusive cross-section, combining \eplus and \muplus
channels, is found to be $\sigma_{\ttbar}$ = 160 $\pm$ 25 pb.
The quoted uncertainty includes both statistical and systematic
contributions and it is dominated by the systematic component. The
result is compatible with the expected \ttbar\ inclusive cross-section
and with previous measurements~\cite{Aad:2012a, Aad:2012b, Chatrchyan:2011a,
  Chatrchyan:2011b}.

\begin{figure*}[htbp]
    \begin{center}
      \subfigure[]{
	\includegraphics[scale=0.37]{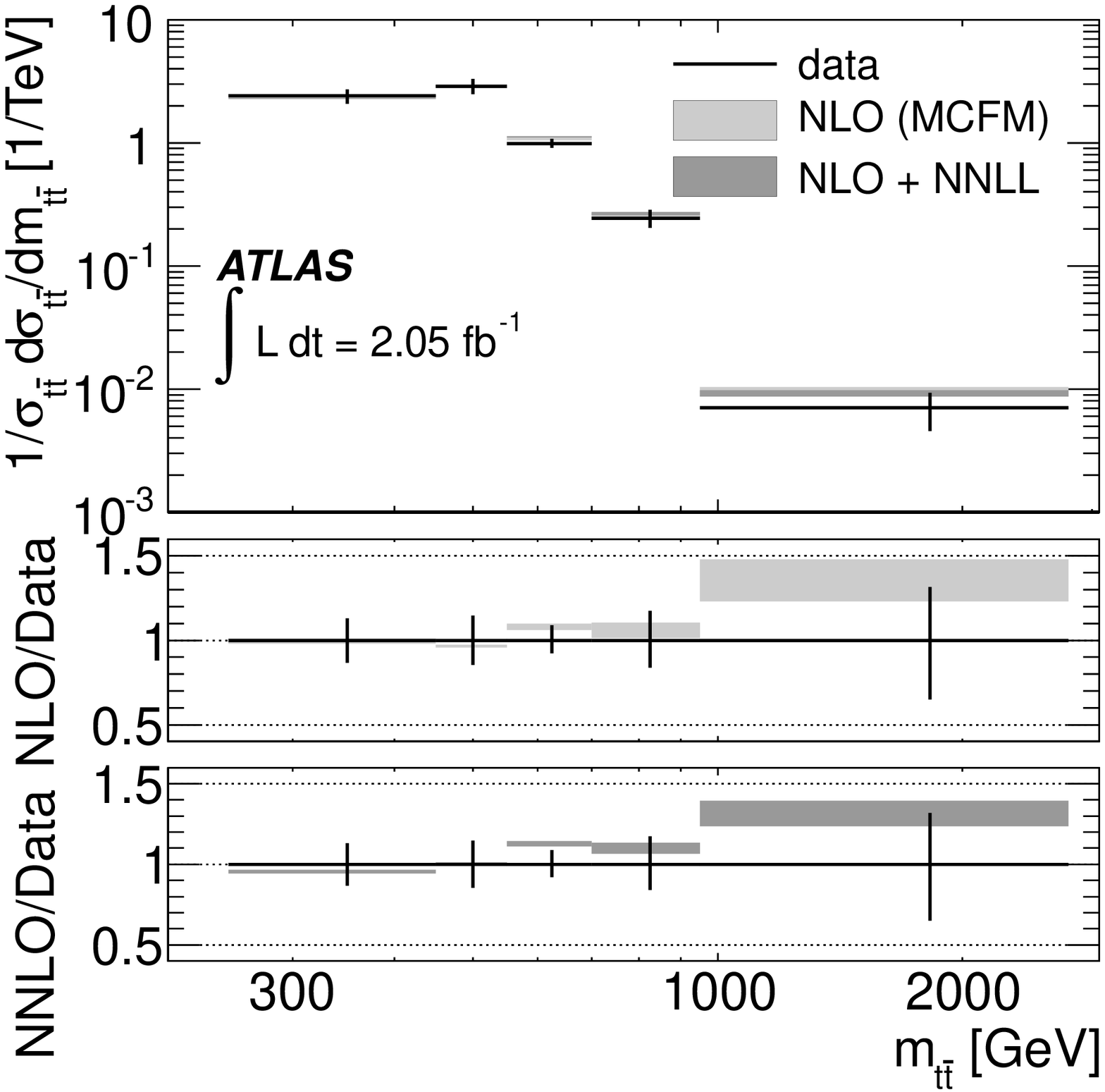}
      }
      \subfigure[]{
	\includegraphics[scale=0.37]{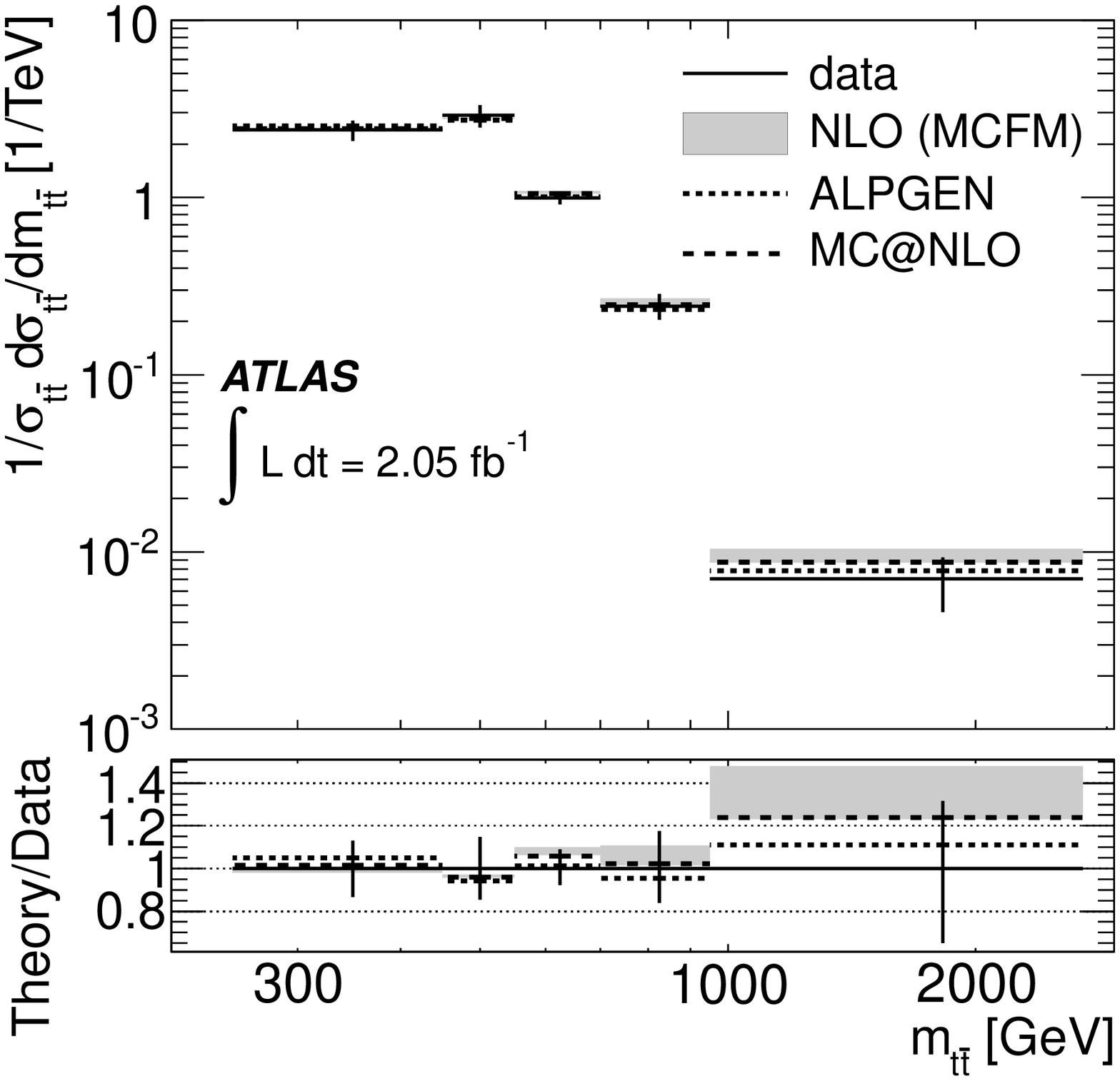}
      }	
      \subfigure[]{
	\includegraphics[scale=0.37]{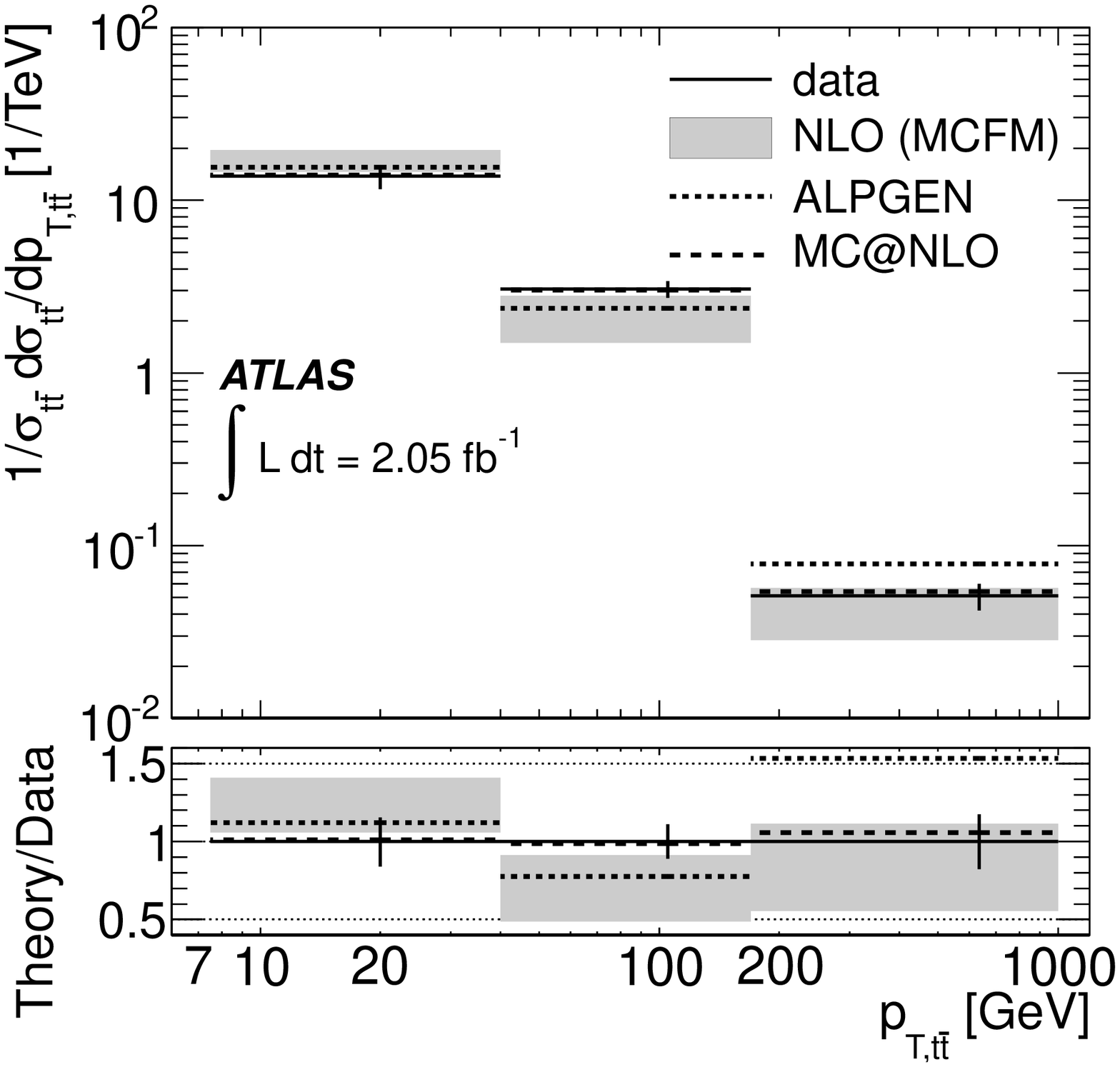}
      }
      \subfigure[]{
	\includegraphics[scale=0.37]{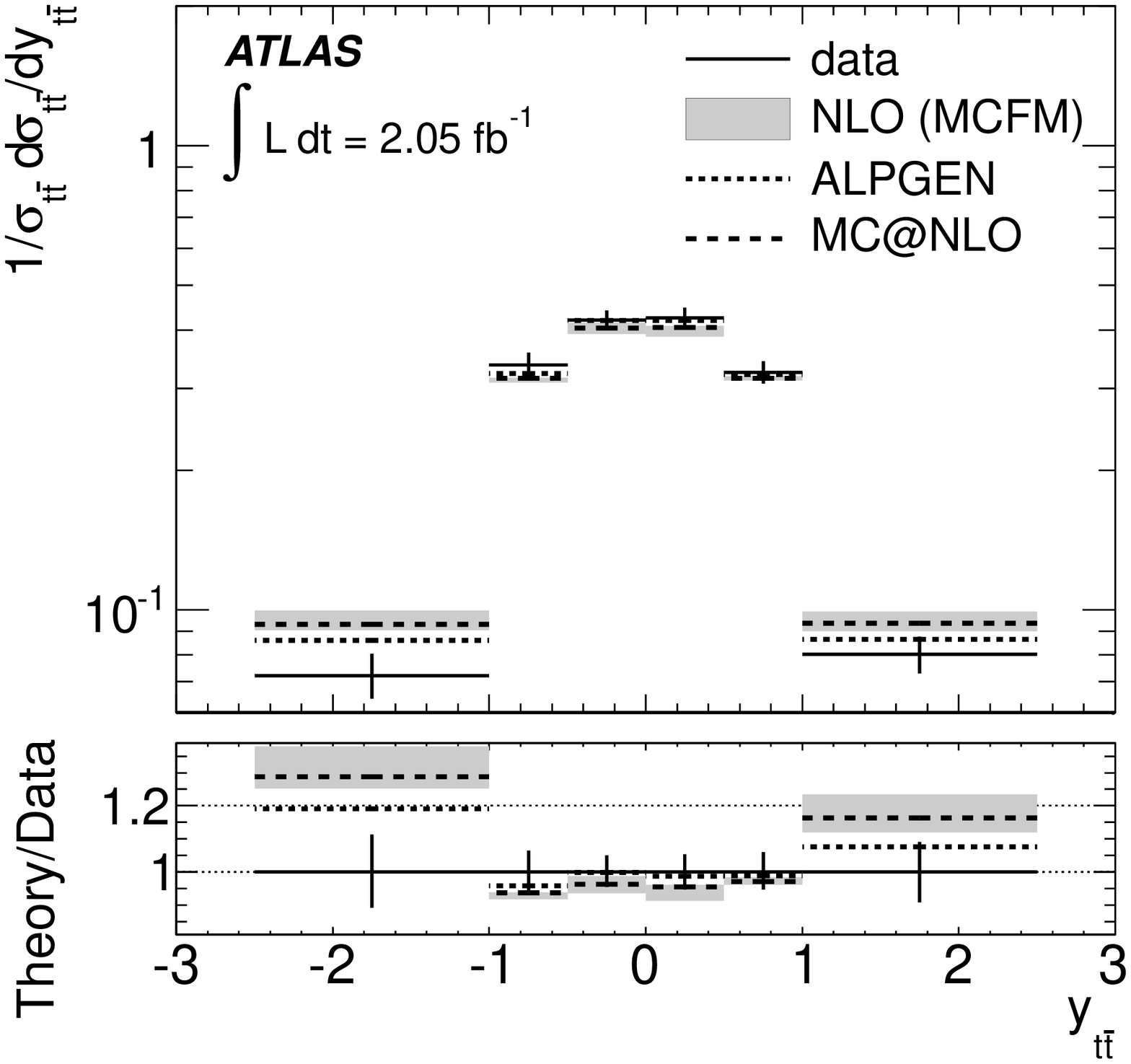}
      }
    \end{center}
  \caption{Relative differential cross-section versus (a-b) $\mttbar$,
    (c) $p_{\mathrm{T},\ttbar}$ and (d) $y_{\ttbar}$.
    Note that the histograms are a graphical
    representation of Table~\ref{tab:DiffXsecMass}.
    This means that only the bin ranges along the $x$-axis (and not
    the position of the vertical error bar) can be
    associated to the relative differential cross-section values on the
    $y$-axis.
    The relative cross-section in
    each bin shown in Table~\ref{tab:DiffXsecMass} is compared
    to the NLO prediction from \mcfm~\cite{MCFM}.
   For $\mttbar$ the results are also compared with the NLO$+$NNLL
   prediction from Ref.~\cite{AHR-0901}.
    The measured uncertainty represents 68\% confidence level
    including both statistical and systematic uncertainties. The
    bands represent theory uncertainties (see Sect.~\ref{sec:results}
    for details).
    Predictions from \mcnlo{} and \alpgen{} are shown for fixed
     settings of the generators' parameters (details are found in
      Sect.~\ref{sec:results}).
    }
  \label{fig:DiffXsecMass}
\end{figure*}

The relative differential cross-section results are listed in
Table~\ref{tab:DiffXsecMass} as a function of  $\mttbar$,
$p_{\mathrm{T},\ttbar}$ and $y_{\ttbar}$. Both
single-channel results and results from the combination are shown.
The correlation coefficients between the measured bins of the combined
result are estimated using
five thousand pseudo-experiments, see Table~\ref{eq:covMatrix}. 
The covariance matrices are derived by combining the correlation coefficients with the
uncertainties  for the respective measurements reported in
Table~\ref{tab:DiffXsecMass} for $\mttbar$,  $p_{\mathrm{T},\ttbar}$ and
$y_{\ttbar}$ respectively. 
A graphical representation for the combined results is shown in
Fig.~\ref{fig:DiffXsecMass}. The measurements are reported with their full
uncertainty, combining statistical and systematic effects, and they are
compared to NLO predictions from \mcfm{}~\cite{MCFM} for all variables;
NLO$+$NNLL predictions from Ref.~\cite{AHR-0901} are included for
$1/\sigma$ $d\sigma/d\mttbar$.
Theory uncertainty bands include uncertainties on parton
distribution functions, the strong coupling constant
$\alpha_S$  and on factorization and renormalization scales.
  For the NLO predictions, the uncertainty from PDFs and $\alpha_S$ is set to the maximal spread of the predictions from three different NLO PDF sets
  (CTEQ6.6, MSTW2008NLO and NNPDF2.0) according to the PDF-specific
  recipe in Ref.~\cite{Botje:2011sn,Nadolsky:2008zw,MSTW2008nlo,NNPDF2010}. Renormalization
  and factorization scales are set to the top quark mass value of 172.5
GeV and associated uncertainties are derived from an upward and downward scale variation of a factor of two.
  The overall NLO uncertainty is obtained by summing the
  contributions from PDFs and  $\alpha_S$  to the contributions from scales in quadrature for variations in the
  same direction.  For the NLO$+$NNLL estimates the uncertainties are
  derived according to the approach of Ref.~\cite{AHR-0901}. The
  uncertainty on the MSTW2008NNLO PDFs and $\alpha_S$  at the 68\% confidence level is combined in quadrature with the
  uncertainties derived from the variations of the factorization scale
  and the renormalization scales.
 For $1/\sigma$ $d\sigma/d\mttbar$ the scale uncertainties are dominant.
    Predictions from \mcnlonobreak{} and
   \alpgen{} are shown for fixed settings of the generators' parameters.
   The settings for \mcnlo{} are given in
    Sect.~\ref{sec:dataSimAndDet}. \alpgen{} is version 2.13 using the
    CTEQ6L1 PDF with the top quark mass set to 172.5 \gev{}. Renormalization
    and factorization scales are set to the same value: the square
    root of the sum of the squared transverse energies of the final state partons. The matching
    parameters~\cite{Mangano:2006rw} for up to five extra partons
    are set to $E_{T}^{clus}$ = 20 \gev{} and $R_{match}$ = 0.7. Parton
    showering and underlying event are simulated by \herwig\ and \jimmy\
    respectively, using the generator tune AUET1~\cite{atlastune}.

\begin{table*}[htbp]
  \caption{Relative differential cross-section
    (top) $1/\sigma$ $d\sigma/d\mttbar$,
    (middle) $1/\sigma$ $d\sigma/d p_{\mathrm{T},\ttbar}$ and (bottom) $1/\sigma$
    $d\sigma/d y_{\ttbar}$ measured in the \eplus,
    \muplus and the combined \lplus channel.}
 \label{tab:DiffXsecMass}
  \begin{center}
    \begin{tabular}{cccc} \hline
      $\mttbar$ [\gev] &
      \multicolumn{3}{c}{ 1/$\sigma$ $d\sigma/d\mttbar$ [1/\tev]} \\  \hline
      & \eplus & \muplus & \lplus \\\hline\hline
      250 -- 450 & 2.2 $\pm$ 0.4 & 2.5 +0.3 / -0.4& 2.4  +0.3 / -0.4\\
      450 -- 550 & 3.3 $\pm$ 0.6 &  2.8 +0.5 / -0.4&2.9 $\pm$ 0.4 \\
      550 -- 700 & 0.9 $\pm $ 0.1 & 1.1 $\pm$ 0.1& 1.0  $\pm$ 0.1 \\
      700 -- 950 & 0.28 $\pm$ 0.06& 0.23  +0.05 / -0.04& 0.24 $\pm$ 0.04\\
      950 -- 2700 & 0.007 $\pm$ 0.003 & 0.008 $\pm$ 0.004 & 0.007 $\pm$ 0.003\\
      \hline\hline
    \end{tabular}

    \vspace{4mm}

    \begin{tabular}{cccc} \hline
      $p_{\mathrm{T},\ttbar}$ [\gev] & \multicolumn{3}{c}{1/$\sigma$
	$d\sigma/d p_{\mathrm{T},\ttbar}$ [1/\tev] } \\  \hline
      & \eplus & \muplus & \lplus \\ \hline\hline
      0 -- 40 &14  $\pm$ 2 &14 $\pm$ 2&14  $\pm$ 2\\
      40 -- 170 &3.0 $\pm$ 0.4 &3.1 $\pm$ 0.3 &3.0 $\pm$ 0.3\\
      170 -- 1100 &0.050 $\pm$ 0.010 &0.051  $\pm$ 0.008 &0.051 $\pm$ 0.008\\
      \hline\hline
    \end{tabular}    

    \vspace{4mm}

    \begin{tabular}{cccc} \hline $y_{\ttbar}$  &
      \multicolumn{3}{c}{1/$\sigma$ $d\sigma/dy_{\ttbar} $ } \\ \hline
      & \eplus & \muplus & \lplus  \\ \hline\hline
      -2.5 -- -1 &0.070 $\pm$ 0.010 & 0.077 $\pm$ 0.009&0.072 $\pm$ 0.008\\
      -1 -- -0.5 &0.32 $\pm$ 0.03& 0.35 $\pm$ 0.03 & 0.34 $\pm$ 0.02\\
      -0.5 -- 0 &0.43 $\pm$ 0.03&0.41 $\pm$ 0.02 & 0.42 $\pm$ 0.02\\
      0 -- 0.5 &0.42 $\pm$ 0.04 &0.43 $\pm$ 0.02 &0.42 $\pm$ 0.02\\
      0.5 -- 1 &0.34 $\pm$ 0.03 &0.31 $\pm$ 0.02 &0.32 $\pm$ 0.02\\
      1 -- 2.5 &0.080  $\pm$ 0.010 &0.083 $\pm$ 0.007 &0.080 $\pm$ 0.007\\
      \hline\hline
    \end{tabular}
  \end{center}
\end{table*}

 \begin{table*}[htbp]
   \caption{Correlation coefficients between bins of the relative
     differential cross-section (top) $1/\sigma$ $d\sigma/d\mttbar$,
     (middle) $1/\sigma$ $d\sigma/d p_{\mathrm{T},\ttbar}$ and
     (bottom) $1/\sigma$
     $d\sigma/d y_{\ttbar}$ in the combined \lplus channel.}
 \label{eq:covMatrix}

$$ C_{\mttbar}  = \begin{bmatrix} 
     1.00&  -0.94&  -0.57&  -0.62&  -0.30\\
     -0.94&  1.00&  0.43&  0.54&  0.20\\
     -0.57&  0.43&  1.00&  0.24&  0.44\\
     -0.62&  0.54&  0.24&  1.00&  0.21\\
     -0.30&  0.20&  0.44&  0.21&  1.00\\
\end{bmatrix} $$

    \vspace{2mm}

$$     C_{p_{\mathrm{T},\ttbar}} = \begin{bmatrix} 
             1.00&  -0.93&  -0.30\\
             -0.93&  1.00&  0.21\\
             -0.30&  0.21&  1.00\\
\end{bmatrix} $$

    \vspace{2mm}
$$         C_{y_{\ttbar}} = \begin{bmatrix} 
             1.00&  -0.61&  0.22&  -0.40&  0.08&  0.21\\
            -0.61&  1.00&  -0.51&  0.24&  -0.13&  0.14\\
            0.22&  -0.51&  1.00&  -0.34&  0.29&  -0.25\\
            -0.40&  0.24&  -0.34&  1.00&  -0.38&  -0.04\\
             0.08&  -0.13&  0.29&  -0.38&  1.00&  -0.58\\
             0.21&  0.14&  -0.25&  -0.04&  -0.58&  1.00\\
\end{bmatrix} $$

\end{table*}

The impact of the different uncertainty
sources on the final results is estimated and shown in
Table~\ref{tab:combDiffXsecSysMass}. 
For $1/\sigma$ $d\sigma/d\mttbar$  the relative
  statistical uncertainty varies from about 2\% at low $\mttbar$ to about 20\% at the
highest $\mttbar$, while the systematic uncertainty ranges between 10\% at
intermediate $\mttbar$ values to about 37\% at the highest $\mttbar$.
In relation to $1/\sigma$ $d\sigma/d p_{\mathrm{T},\ttbar}$ the relative statistical uncertainty
ranges between about 4\% at low $p_{\mathrm{T},\ttbar}$  and about 12\% at the
highest $p_{\mathrm{T},\ttbar}$ values, while the systematic
uncertainty increases from about 13\%  to 20\%  in the same interval.
In the case of $1/\sigma$ $d\sigma/d y_{\ttbar} $ the relative statistical
uncertainty increases from about 3\%  at low $y_{\ttbar}$ to about 5\% at the
highest $y_{\ttbar}$ values, while the systematic uncertainty changes 
from 4\% to 10\%  over the same interval.
 Jet-related uncertainties are  dominant for $\mttbar$ and $p_{\mathrm{T},\ttbar}$, while for
$y_{\ttbar}$ the dominant contributions are from fake-leptons and final-state
radiation in addition to the jet uncertainties.

No significant deviations from the SM expectations provided by the
different MC generators are observed. The SM prediction for the relative cross-section distribution can be
tested against the measured values by using the covariance matrix
between the measured bins of the combined results.

\section{Conclusions}
\label{sec:conclusions}

Using a dataset of 2.05 \invfb, the relative differential cross-section
for \ttbar\ production is  measured as a function of three properties
of the \ttbar\ system: mass ($\mttbar$),  \pt\ ($p_{\mathrm{T},\ttbar}$)
and rapidity
($y_{\ttbar}$). The background-subtracted, detector-unfolded values of
$1/\sigma$ $d\sigma/d\mttbar$, $1/\sigma$ $d\sigma/d p_{\mathrm{T},\ttbar}$
and $1/\sigma$ $d\sigma/d y_{\ttbar} $ are reported together with
their respective covariance matrices, and compared to theoretical calculations.
The measurement uncertainties range typically between 10\% and 20\%
and are generally dominated by systematic effects.
No significant deviations from the SM expectations provided by the
different MC generators are observed.

\begin{table*}[tbhp]
  \caption{Percentage uncertainties on (top) $1/\sigma$ $d\sigma/d\mttbar$,
    (middle) $1/\sigma$ $d\sigma/d p_{\mathrm{T},\ttbar}$ and
    (bottom) $1/\sigma$ $d\sigma/d
    y_{\ttbar}$ in the combined \lplus channel.}
  \label{tab:combDiffXsecSysMass}
  \begin{center}
    {\footnotesize
      \begin{tabular}{lccccc} \hline
	$1/\sigma$ $d\sigma/d\mttbar$ & \multicolumn{5}{c}{ $\mttbar$ bins [\gev]}\\
	\hline 
	Uncertainty [$\%$] &250 -- 450  & 450 -- 550 & 550 -- 700& 700 -- 950 & 950 -- 2700\\ \hline\hline
	Total &14 / -14&15 / -15&10 / -10&18 / -16&37 / -43\\
	Stat only &2 / -2&4 / -4&5 / -5&8 / -8&18 / -19\\
	Syst. only &14 / -14&14 / -15&8 / -8&16 / -14&32 / -37\\ \hline
	Luminosity &1 / -1&2 / -2&2 / -1&1 / -1&1 / -2\\
	Jets&11 / -10&10 / -11&6 / -6&13 / -11&20 / -24\\
	Leptons &1 / -1&1 / -1&1 / -2&2 / -2&9 / -6\\
	\MET energy scale &1 / -1&1 / -1&1 / -2&2 / -1&9 / -5\\
	Fake-lepton and $W$ backgrounds &5 / -7&10 / -7&5 / -4&5 / -6&10 / -15\\
  Monte Carlo gen., theory, ISR/FSR, and PDF &6 /-7&7/-7&4/-4&8/-7&14/-18\\
	\hline\hline
      \end{tabular}
    }
   
    \vspace{4mm}

    \begin{tabular}{lccc} \hline
      $1/\sigma$ $d\sigma/d p_{\mathrm{T},\ttbar}$&
      \multicolumn{3}{c}{$p_{\mathrm{T},\ttbar}$ bins [\gev] }\\  \hline 
      Uncertainty [$\%$]  &0 -- 40  & 40 -- 170 & 170 -- 1100\\ \hline\hline
      Total &14 / -16&13 / -12&23 / -22\\
      Stat. only &4 / -4&4 / -5&12 / -11\\
      Syst. only &13 / -16&12 / -11&20 / -19\\ \hline
      Luminosity &1 / -1&2 / -2&2 / -5\\
      Jets &8 / -7&6 / -7&11 / -10\\
      Leptons &1 / -1&1 / -1&2 / -2\\
      \MET energy scale &4 / -4&4 / -4&3 / -1\\
      Fake-lepton and $W$ backgrounds &2 / -5&5 / -3&7 / -4\\
      Monte Carlo gen., theory, ISR/FSR, and PDF &10 / -13&6 / -6&8 / -7\\
      \hline \hline
    \end{tabular}
   
    \vspace{4mm}
    
    \begin{tabular}{lcccccc} \hline
      $1/\sigma$ $d\sigma/d y_{\ttbar}$ & \multicolumn{6}{c}{ $y_{\ttbar}$ bins}\\ 
      \hline  
      Uncertainty [$\%$]      &-2.5 --  -1&-1 -- -0.5 & -0.5 -- 0&0 -- 0.5 & 0.5 -- 1  &1 -- 2.5\\ \hline\hline
      Total&11 / -10&7 / -7&5 / -5&5 / -5&6 / -5&9 / -9\\
      Stat. only &5 / -5&4 / -4&3 / -3&3 / -4&4 / -4&5 / -5\\
      Syst. only &10 / -9&5 / -5&4 / -3&4 / -4&4 / -3&7 / -7\\ \hline
      Luminosity &1 / -2&1 / -1&1 / -1&1 / -1&1 / -1&1 / -1\\
      Jets &4 / -4&1 / -1&1 / -1&2 / -2&1 / -1&3 / -3\\
      Leptons &1 / -1&1 / -1&1 / -1&1 / -1&1 / -1&1 / -2\\
      \MET energy scale &1 / -2&1 / -2&1 / -1&1 / -1&1 / -1&1 / -1\\
      Fake-lepton and $W$ backgrounds &4 / -7&4 / -2&1 / -1&1/-1&1 / -1&1 / -3\\
      Monte Carlo gen., theory, ISR/FSR, and PDF &6 / -5&3 / -4&3 / -3&2 /-2&3/-2&4/-6\\
      \hline\hline
    \end{tabular}
 \end{center}
\end{table*}

\section{Acknowledgements}

We thank CERN for the very successful operation of the LHC, as well as the
support staff from our institutions without whom ATLAS could not be
operated efficiently.

We acknowledge the support of ANPCyT, Argentina; YerPhI, Armenia; ARC,
Australia; BMWF and FWF, Austria; ANAS, Azerbaijan; SSTC, Belarus; CNPq and FAPESP,
Brazil; NSERC, NRC and CFI, Canada; CERN; CONICYT, Chile; CAS, MOST and NSFC,
China; COLCIENCIAS, Colombia; MSMT CR, MPO CR and VSC CR, Czech Republic;
DNRF, DNSRC and Lundbeck Foundation, Denmark; EPLANET, ERC and NSRF, European Union;
IN2P3-CNRS, CEA-DSM/IRFU, France; GNSF, Georgia; BMBF, DFG, HGF, MPG and AvH
Foundation, Germany; GSRT and NSRF, Greece; ISF, MINERVA, GIF, DIP and Benoziyo Center,
Israel; INFN, Italy; MEXT and JSPS, Japan; CNRST, Morocco; FOM and NWO,
Netherlands; BRF and RCN, Norway; MNiSW, Poland; GRICES and FCT, Portugal; MERYS
(MECTS), Romania; MES of Russia and ROSATOM, Russian Federation; JINR; MSTD,
Serbia; MSSR, Slovakia; ARRS and MVZT, Slovenia; DST/NRF, South Africa;
MICINN, Spain; SRC and Wallenberg Foundation, Sweden; SER, SNSF and Cantons of
Bern and Geneva, Switzerland; NSC, Taiwan; TAEK, Turkey; STFC, the Royal
Society and Leverhulme Trust, United Kingdom; DOE and NSF, United States of
America.
\newline
\indent The crucial computing support from all WLCG partners is acknowledged
gratefully, in particular from CERN and the ATLAS Tier-1 facilities at
TRIUMF (Canada), NDGF (Denmark, Norway, Sweden), CC-IN2P3 (France),
KIT/GridKA (Germany), INFN-CNAF (Italy), NL-T1 (Netherlands), PIC (Spain),
ASGC (Taiwan), RAL (UK) and BNL (USA) and in the Tier-2 facilities
worldwide.

\bibliographystyle{atlasnote}
\bibliography{TopDiffXsec}
\clearpage
\begin{flushleft}
{\Large The ATLAS Collaboration}

\bigskip

G.~Aad$^{\rm 47}$,
T.~Abajyan$^{\rm 20}$,
B.~Abbott$^{\rm 110}$,
J.~Abdallah$^{\rm 11}$,
S.~Abdel~Khalek$^{\rm 114}$,
A.A.~Abdelalim$^{\rm 48}$,
O.~Abdinov$^{\rm 10}$,
R.~Aben$^{\rm 104}$,
B.~Abi$^{\rm 111}$,
M.~Abolins$^{\rm 87}$
% ...
\bigskip
\\
% ...
$^{10}$ Institute of Physics, Azerbaijan Academy of Sciences, Baku, Azerbaijan\\
$^{11}$ Institut de F{\'\i}sica d'Altes Energies and Departament de F{\'\i}sica de la Universitat Aut{\`o}noma de Barcelona and ICREA, Barcelona, Spain\\
% ...
$^{20}$ Physikalisches Institut, University of Bonn, Bonn, Germany\\
% ...
$^{47}$ Fakult{\"a}t f{\"u}r Mathematik und Physik, Albert-Ludwigs-Universit{\"a}t, Freiburg, Germany\\
$^{48}$ Section de Physique, Universit{\'e} de Gen{\`e}ve, Geneva, Switzerland\\
% ...
$^{87}$ Department of Physics and Astronomy, Michigan State University, East Lansing MI, United States of America\\
% ...
$^{104}$ Nikhef National Institute for Subatomic Physics and University of Amsterdam, Amsterdam, Netherlands\\
% ...
$^{110}$ Homer L. Dodge Department of Physics and Astronomy, University of Oklahoma, Norman OK, United States of America\\
$^{111}$ Department of Physics, Oklahoma State University, Stillwater OK, United States of America\\
% ...
$^{114}$ LAL, Universit{\'e} Paris-Sud and CNRS/IN2P3, Orsay, France\\
% ...
\end{flushleft}

 
\end{document}